\documentclass[12,Arial]{article}
\usepackage[margin=1in]{geometry}
\usepackage{amsmath}
\usepackage{array}
\usepackage{graphicx}
\usepackage{subcaption}
\usepackage{color,soul}
\usepackage[hang,flushmargin]{footmisc}
\usepackage[colorlinks=true, allcolors=blue]{hyperref}
\usepackage{setspace}
\usepackage{booktabs}
\usepackage{tikz}
\usepackage{longtable}
\usepackage{pdflscape}
\usepackage{multirow}

\usetikzlibrary {shapes.geometric}
\usepackage[utf8]{inputenc}
\usepackage[english]{babel}

\usepackage{natbib}
\setcitestyle{authoryear,open={(},close={)}} 

\title{Watts and Bots: The Energy Implications of AI Adoption}
\author{ Anthony R. Harding \\ Georgia Institute of Technology\\tony.harding@gatech.edu \and Juan Moreno-Cruz \\ University of Waterloo\\jmorenoc@uwaterloo.ca
}
\date{\today}

\begin{document}
\maketitle 
\begin{abstract}
    With the rapid expansion of Artificial Intelligence, there are expectations for a proportional expansion of economic activity due to increased productivity, and with it energy consumption and its associated environmental consequences like carbon dioxide emissions. Here, we combine data on economic activity, with early estimates of likely adoption of AI across occupations and industries, to estimate the increase in energy use and carbon dioxide emissions at the industry level and in aggregate for the US economy. At the industry level, energy use can increase between 0 and 12 PJ per year, while emissions increase between 47 tCO$_2$ and 272 ktCO$_2$. Aggregating across industries in the US economy, this totals an increase in energy consumption of 28 PJ per year, or around 0.03\% of energy use per year in the US. We find this translates to an increase in carbon dioxide emissions of 896 ktCO$_2$ per year, or around 0.02\% of the CO$_2$ emissions per year in the US.
\end{abstract}

\vfill
\noindent {\small \textbf{Funding Disclosure:} ARH and JMC were supported by funding from Google. The funder had no role in the design, analysis, interpretation, or writing of this paper. All findings and conclusions are solely those of the authors.}

\clearpage

\section{Introduction}

Rapid advancement and adoption of artificial intelligence (AI) technologies have promised unprecedented productivity gains across the economy. As of February 2024 around 5.4\% of firms were using AI, up 1.7 percentage points from 5 months earlier and expected to rise another 1.2 percentage points over the following 6 months \citep{bonney_tracking_2024}. The promise of AI comes with energy and environmental challenges \citep{luers_will_2024}.

AI technologies require significant energy. This energy is associated with data storage, cooling systems, and the production of specialized hardware \citep{de_vries_growing_2023}. For example, training large AI models can consume as much energy as several hundred households in a year and the energy use of these models in inference can be considerably higher \citep{strubell_energy_2020,de_vries_growing_2023}. Yet, the effects of AI on energy consumption extend beyond direct electricity use for computing.  Energy is a critical input in almost all economic activities, powering industries, enabling transportation, and supporting modern life's infrastructure. Many studies have shown a strong correlation between energy consumption and economic output, demonstrating that energy use is closely linked to GDP growth \citep{ayres_economic_2010}. There is potential for AI to reduce energy use and expedite the decarbonization of the economy by improving energy efficiency of production processes (e.g. \citet{john_how_2022}), improving the efficiency of energy systems (e.g. \citet{noori_artificial_2024}) or through demand-side management (e.g. \citet{antonopoulos_artificial_2020}). However, the predominant reliance on fossil fuels for energy generation result in increased air pollution \citep{oecd2016energy}, degraded water quality, land-use and climate change \citep{howells2013integrated}. If AI increases economic productivity, it may increase aggregate energy use and amplify existing environmental impacts of production.

We provide the first quantitative estimate of the impact of AI adoption on energy use and carbon emissions at both the industry and aggregate levels for the U.S. economy. We develop a parsimonious model to quantify the change in total energy use and emissions due to the adoption of AI by firms for production. Our model and estimates provide insights into its potential future impact, highlighting the importance of variation across economic industries in the economy in terms of their likelihood to benefit from the adoption of AI technologies as well as their energy and emissions intensities. This supports better design of policies that balance the benefits of AI advancement with energy and emissions sustainability goals.

\section{Methods}
\begin{figure}[h!]
\centering
\begin{subfigure}[t]{0.67\textwidth}
        \centering
        \includegraphics[width=\linewidth]{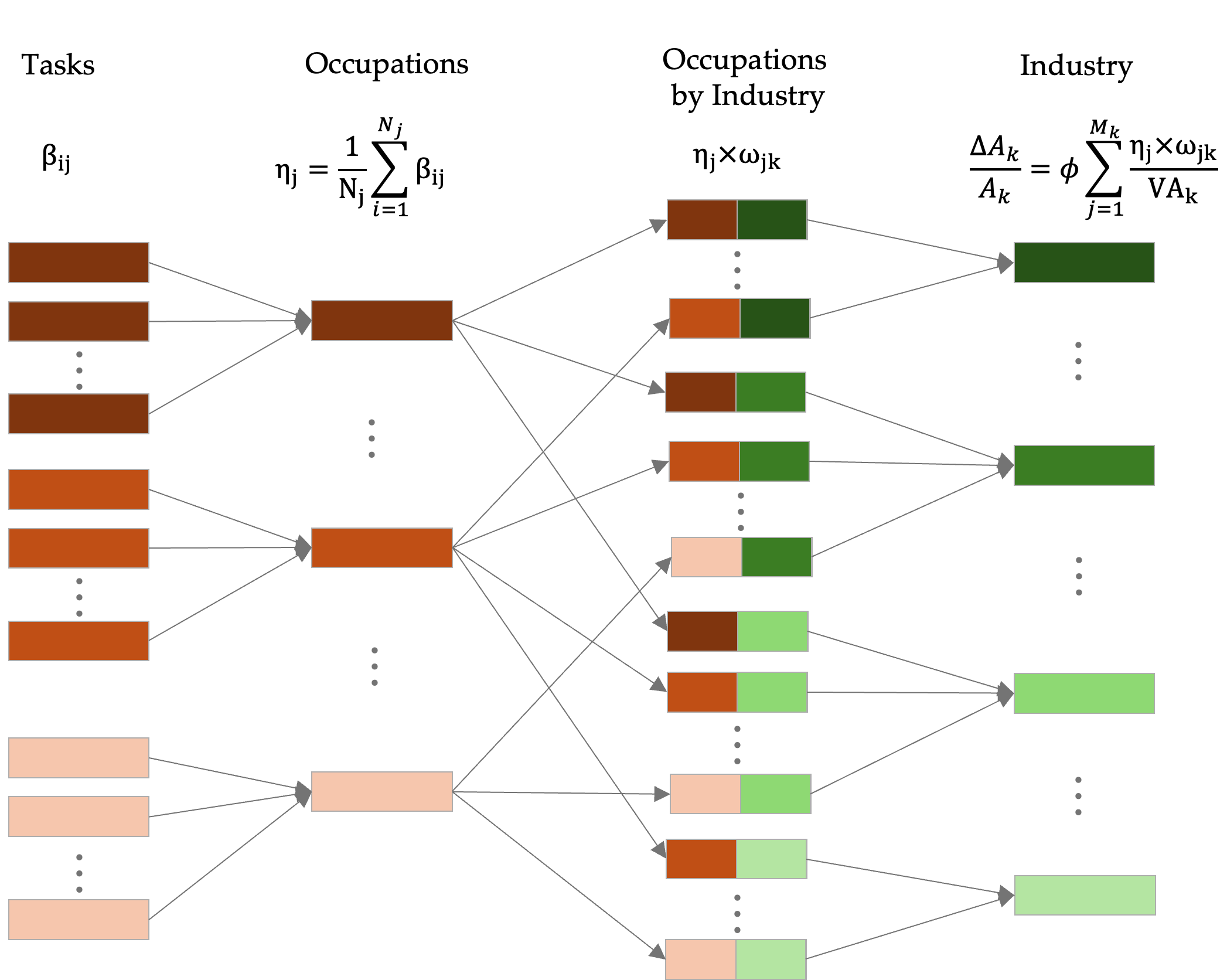}
        \caption{Aggregation Process}
    \end{subfigure}%
   ~
    \begin{subfigure}[t]{0.33\textwidth}
        \centering
        \includegraphics[width=\linewidth]{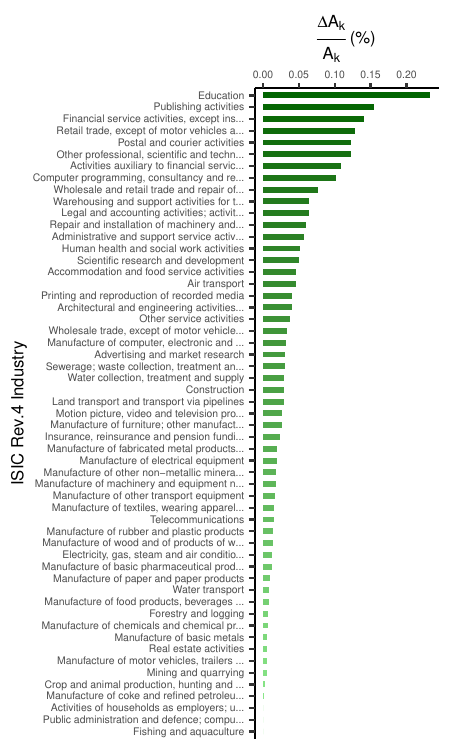}
        \caption{Productivity shock}
    \end{subfigure}
    \caption{Exposure to AI by Industry}
    \label{fig:Domar}
\end{figure}

We propose a first-order approximation of the energy and environmental impacts of AI use in the economy, using a theoretically founded partial equilibrium model of the US economy \cite{hulten_growth_1978} (See Supplementary Materials \ref{app:Theoretical framework}). Figure \ref{fig:Domar} illustrates our approach to aggregate from highly specific task-level AI exposure data to broad industry-wide impacts (See Detailed methods in Supplementary Materials \ref{sec:methods}). By aggregating from tasks to occupations and then to industries, we can account for the heterogeneous influence of AI across different sectors of the economy. 

We begin with the exposure of tasks--specific activities or units of work performed within a job--to AI, $\beta_{ij}$. Exposure scores for 19,265 tasks come from expert survey data from \citet{eloundou_gpts_2024}. AI exposure of each occupation is measured as the average of the AI exposure of its tasks. Occupational AI exposure is aggregated to the industry-level, measuring AI exposure for each of 55 industries as the average exposure of occupations within the industry weighted by their wage-bill share \citep{us_bureau_of_labor_statistics_last_occupational_2024}. Finally, these AI exposure measures are merged with economic, energy, and emissions data for the US using the World Input-Output Database (WIOD) 2016 release and corresponding environmental accounts \citep{timmer_illustrated_2015,corsatea_world_2019}. This multi-step process allows us to consider not only which tasks are susceptible to AI, but also how important those tasks are within occupations, and how significant those occupations are within industries. It provides a comprehensive view of AI's potential impact across the economy, linking granular task-level data to broad industry-wide effects.

We follow the approach of \citet{acemoglu_simple_2024} to translate AI exposure to cost savings using a uniform scaling Cost Savings factor, $\phi<1$, that is the product of two parameters: the average fraction of exposed tasks that are profitably replaced by AI and the average fraction of labor costs saved for a task if AI is adopted. We follow \citet{acemoglu_simple_2024} estimates that 23\% of exposed tasks replace human labor with AI and that replacing human labor with AI results in 27\% cost savings. This gives a Cost Savings factor of $\phi=0.27 \times 0.23 = 0.0621$. Given their importance, we illustrate the sensitivity of our estimates to these parameters in Figure \ref{fig:sensitivity}.

We can then write the percentage change in productivity as
\begin{equation}\label{eq:task impact}
    \frac{\Delta y_k}{y_k}=\frac{\Delta A_k}{A_k} =\phi\sum_j^{M_k}\frac{\eta_{j}\times \omega_{jk}}{y_k}
\end{equation}
where the first identify follows from assuming that percentage changes in economic output are proportional to percentage changes in the productivity of the economy. That is, productivity affects all production inputs in the same proportion. 

Putting this together, the right-hand panel of Figure \ref{fig:Domar} shows estimates of the productivity impact of AI exposure across industries (See Table \ref{tab:industry_changes}). We estimate a range of productivity impacts of AI from 0.233\% for the education industry to 0\% for industries like fishing and aquaculture. 

To translate the economic productivity impacts of AI exposure to changes in energy use and carbon emissions we examine the relationships between industry-level productivity, energy intensity $\nu_k$, and carbon intensity $\mu_k$. Specifically we calculate the change in economic output, the change in energy, and the change in CO$_2$ emissions at the industry level as
\begin{align}\label{eq:industry impact}
    \Delta y_k&=y_k \Delta A_k/A_k\\
    \Delta E_k& = \nu_{k} \Delta y_k \\
    \Delta C_k& = \mu_{k} \Delta E_k
\end{align}

Changes in output due to AI-driven productivity lead to a proportional increase in energy use, scaled by the industry's energy intensity. And changes in energy use lead to a proportional increase in carbon emissions, scaled by the industry's emissions intensity. Together, these equations allow us to trace the impact of AI-driven productivity changes on energy use and carbon emissions, accounting for the specific characteristics of each industry.

\section{Results}

\begin{figure}[h!]
\centering
\begin{subfigure}[t]{0.33\textwidth}
        \centering
        \includegraphics[width=\linewidth]{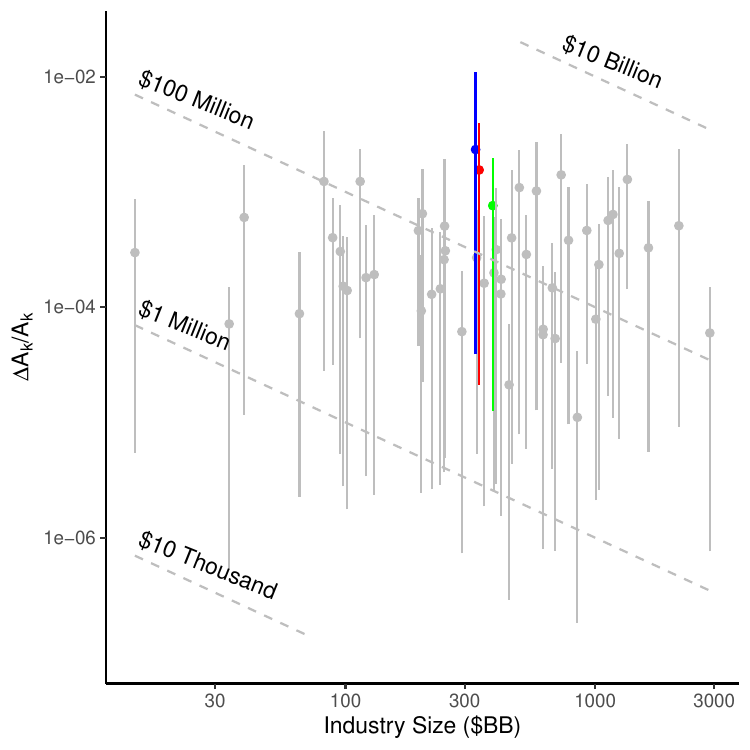}
        \caption{Change in GDP}
    \end{subfigure}%
    ~ 
    \begin{subfigure}[t]{0.33\textwidth}
        \centering
        \includegraphics[width=\linewidth]{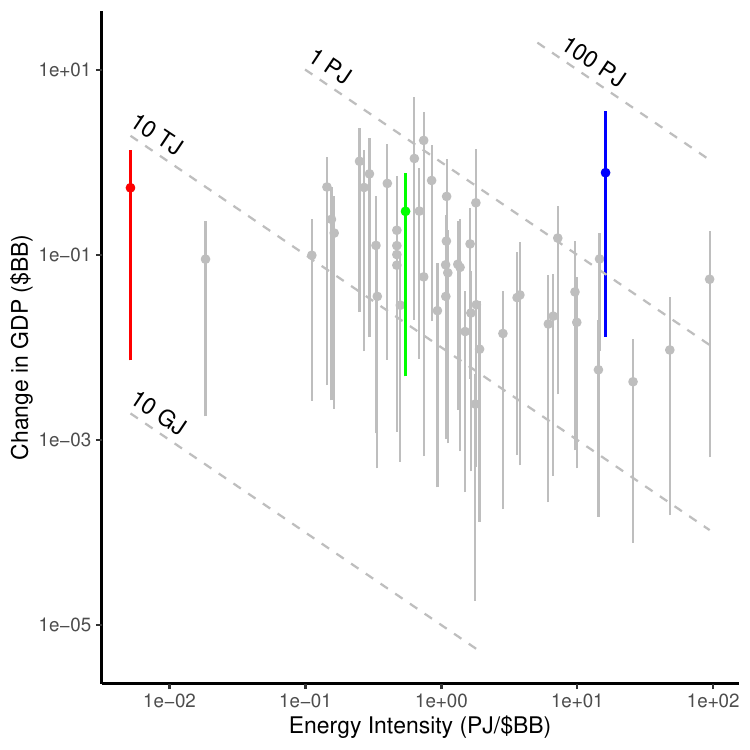}
        \caption{Change in Energy Use }
    \end{subfigure}%
    ~
    \begin{subfigure}[t]{0.33\textwidth}
        \centering
        \includegraphics[width=\linewidth]{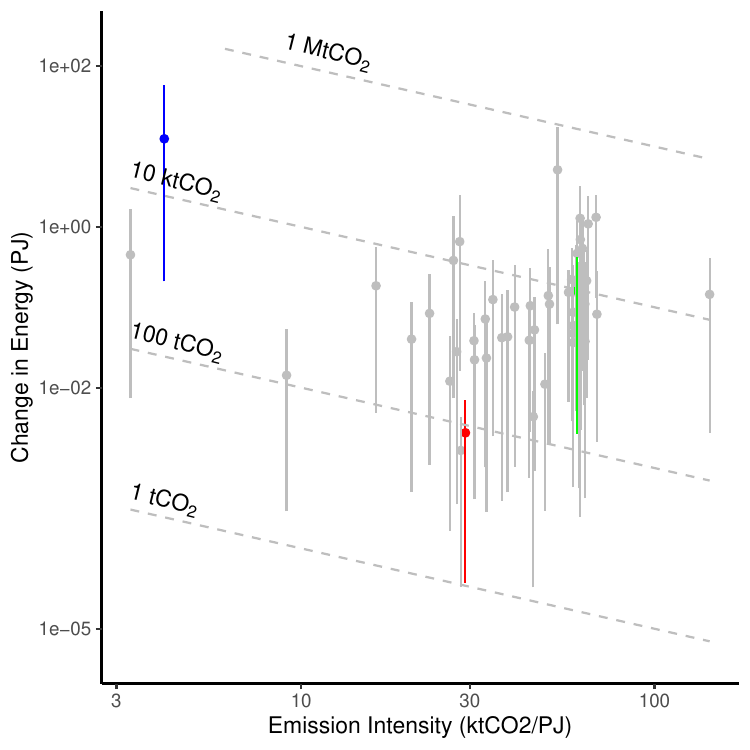}
        \caption{Change in Carbon Emissions}
    \end{subfigure}
    \caption{The Impact of AI Adoption on Output (a), Energy (b), and CO$_2$ Emissions (c). Each point and bar represents the points showing our central estimates and the bars showing the range of estimates across different assumptions regarding AI penetration rates from estimates in  \citet{eloundou_gpts_2024}. In color are ISIC Rev. 4 Code P85: {\color{blue} Education (Blue)}, ISIC Rev. 4 Code J58: {\color{red} Publishing Activities (Red)}, and ISIC Rev. 4 Code G45: {\color{green} Wholesale and retail trade and repair of motor vehicles and motorcycles (Green).}}
    \label{fig:impact_industry}
\end{figure}

Figure \ref{fig:impact_industry} presents our main results for changes in output, energy use, and carbon emissions, corresponding to Equations (2), (3), and (4). 

On average, an industry in our study experiences a change in energy use of 0.511 PJ and a change in carbon emissions of 16 ktCO$_2$. This average masks significant variations across industries. And our estimates do not reveal a clear relationship between industry size and the magnitude of the AI-induced shock. The spread of impacts can be substantial, varying across orders-of-magnitude. Due to our models proportionality relationships, smaller industries generally experience smaller absolute changes in output, less energy intense industries generally experience smaller changes in energy use, and less emissions intense industries experience smaller changes in carbon emissions. For example, in Table \ref{tab:example_industries} we summarize the impacts for three distinct industries. Initially, these industries are comparable in terms of the productivity impact of AI adoption and their economic size. Thus, they are in a similar vertical position in Figure 2b. However, they have a large spread in energy use, leading to a large spread in the energy use impacts of AI adoption. And, there is a similarly large spread in the change in carbon emissions.

\begin{table}[h!]
\caption{The Impact of AI on Output, Energy, and CO$_2$ Emissions: Selected Industries}
\label{tab:example_industries}
\centering
\scalebox{0.75}{
\begin{tabular}{l|cccc|ccc}
\hline
 \multirow{2}{*}{\textbf{Industry}} & \textbf{Output}  & \textbf{Exposure rates} &\textbf{Energy Intensity} & \textbf{Emissions Intensity}  & $\boldsymbol{\Delta y_k}$ & $\boldsymbol{\Delta E_k}$ & $\boldsymbol{\Delta C_k}$ \\
  & (\$BB) & (\%) &(PJ/\$BB) & (ktCO$_2$/PJ)  & (\$BB) & (PJ) & (ktCO$_2$) \\\hline
 {\color{blue}Education} &   332 &0.233   & 16.13 &4.10 &0.774 & 12.477 & 51.133\\
{\color{red}Publishing Activities}         &   343    & 0.155&0.01& 29.15 & 0.531 & 0.003 &  0.08 \\
{\color{green}Trade and Repair of motor vehicles}         &   389    &0.076 &0.54 & 60.29 & 0.296& 0.161 & 9.711    \\\hline
\end{tabular}
}
\end{table}

Change in energy use, driven by industry-specific factors, largely determines the change in emissions. The education sector, for instance, shows a large increase in both energy use and emissions. This could be due to factors such as the energy-intensive nature of educational infrastructure or the widespread application of AI across various educational processes. 

These results highlight the importance of considering industry-specific characteristics when assessing the energy and environmental impacts of AI adoption. Factors such as the current energy efficiency of an industry, the nature of its processes, and its capacity to integrate AI technologies all play important roles in determining the ultimate energy and emissions outcomes.

\subsection{Aggregation}

Aggregating across the 55 industry-specific impacts, the total annual change in energy use in the economy and the increase in carbon dioxide emissions are given by
\begin{align}\label{eq:Aggreagte Impacts}
    \Delta E&=\sum_{k=1}^{55} \Delta E_k=28 PJ\\
    \Delta C&=\sum_{k=1}^{55} \Delta C_k = 897 ktCO_2
\end{align}
As an aggregate change due to productivity gains from AI adoption, this first-order approximation encompasses the change in energy use across the whole economy, not just the energy required for training and running AI models. 

So how does this compare to bottom-up estimates of the direct energy use of AI? For model training, the energy use of four widely used Large Language Models (LLMs)—BLOOM, GPT-3, Gopher, and OPT—ranged from 324 MWh to 1,287 MWh \citep{de_vries_growing_2023}. However, this is likely small compared to the ongoing energy expenditure for model inference. For instance, ChatGPT inference alone requires around 564 MWh per day or 0.2 Twh per year \citep{de_vries_growing_2023}. Yet, our estimate of 28 PJ (equivalent to about 7.8 TWh) per year is approximately 38 times larger than the annual energy use for ChatGPT inference.

We can also consider hardware-based estimates. NVIDIA, which holds a 95\% market share, was expected to deliver 100,000 AI servers in 2023. de Vries estimates that these servers, running at full capacity, consume approximately 5.7-8.9 TWh of electricity annually. Our estimate of 28 PJ (equivalent to about 7.8 TWh) suggests that the aggregate energy use from AI adoption is comparable to the direct electricity use of new hardware if run at full capacity, ranging from 88\% to 137\%. This indicates close alignment between our top-down and existing bottom-up estimates.

What does this change in energy use mean for generation capacity? An additional 28 PJ (or 7.8 TWh) per year implies that the US would require approximately 0.9 GW of additional generation capacity annually to keep up with the rising demand due to AI adoption-related productivity gains. This represents about 0.08\% of the total U.S. electricity generating capacity in 2021 (1,144 GW), indicating a relatively small but non-negligible impact on the national power infrastructure.

An increase of 897 ktCO$_2$ emissions per year represents less than 1\% of CO$_2$ emissions from the US manufacturing and construction industries and is comparable to the annual emissions of a small country, like Iceland. In the context of the US, it accounts for approximately 0.02\% of total CO$_2$ emissions in 2021 (about 5 GtCO$_2$), indicating that while the impact is measurable, it's a small fraction of overall emissions.

\subsection{Sensitivity Analysis}
\begin{figure}[h!]
\centering
    \begin{subfigure}[t]{0.5\textwidth}
        \centering
        \includegraphics[width=\linewidth]{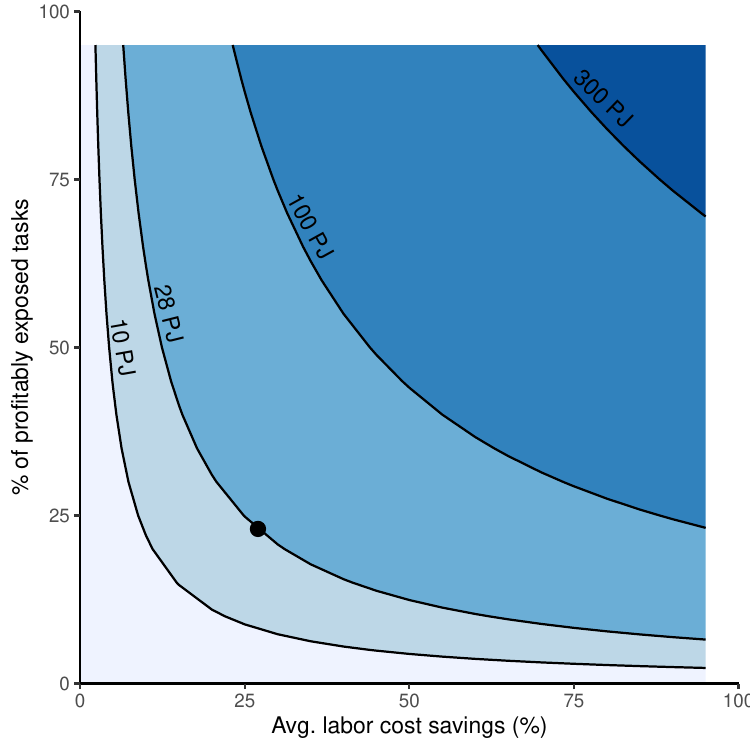}
        \caption{Change in Energy Use }
    \end{subfigure}%
    ~
    \begin{subfigure}[t]{0.5\textwidth}
        \centering
        \includegraphics[width=\linewidth]{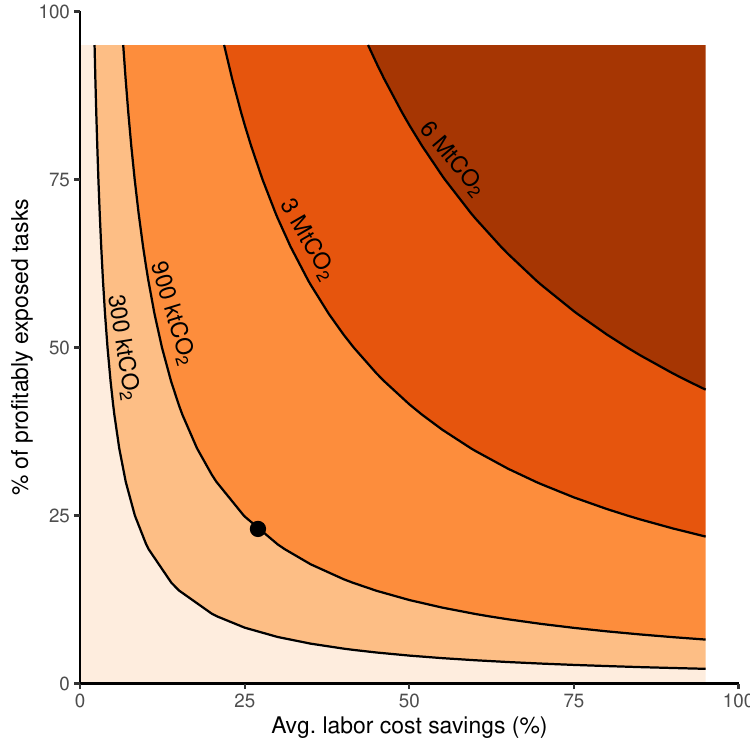}
        \caption{Change in Carbon Emissions}
    \end{subfigure}
    \caption{Sensitivity to cost savings}
    \label{fig:sensitivity}
\end{figure}

Our analysis above relies on a single estimate of the Cost Savings factor derived from limited existing literature. Recognizing the uncertainty surrounding this crucial parameter, in Figure \ref{fig:sensitivity} we perform a sensitivity analysis to set bounds on how variations in the Cost Savings factor affect our estimates.

We illustrate the sensitivity of our estimates of the change in energy use (\ref{fig:sensitivity}(a)) and change in carbon emissions (\ref{fig:sensitivity}(b)) to the two measures that determine the Cost Savings factor--the fraction of tasks that are profitably exposed to AI and the average labor cost savings for tasks performed by AI. Changes in both energy use and carbon emissions are bounded by the contribution of each industry to total economic output. This means that even in the most optimistic AI adoption scenarios---where cost savings and task exposure are at their highest---the changes in energy use and carbon emissions are still limited to a few percentage points of current levels.

These findings suggest that, while AI adoption may lead to increases in energy use and emissions, these increases are unlikely to be dramatically large relative to the overall economy, even under highly favorable conditions for AI.  The sensitivity analysis also shows a non-linear relationship between cost savings, task exposure, and environmental impacts. Small changes in these parameters can lead to large changes in energy use and emissions, particularly in the middle ranges of the parameters. This underscores the importance of accurate estimates for these parameters and highlights areas where further research could significantly improve our understanding of AI's energy and environmental impacts.

\section{Discussion}
Our analysis provides a first (and first order) estimate of the impact of AI adoption on energy use and carbon emissions across the US economy. Here we highlight the assumptions and limitations inherent in our approach. We attempt to summarize and sign the expected effect on our estimates in Table \ref{tab:limitations}, distinguishing between the limitations of the modeling approach and those of the available data. 

\begin{table}[h!]
\caption{The (plausible) effect of assumptions and data limitations on our estimates}\label{tab:limitations}
\begin{tabular}{lc}
 & Overestimate (+)/Underestimate (-) \\\hline
 \textbf{Model Assumptions/Limitations} & \\\hline
AI only affects productivity through labor       &       -    \\
AI does not affect energy efficiency          &       +  \\
Prices and factors of production are fixed & +/-     \\
No changes in the composition of the economy & +/- \\\hline
\textbf{Data Limitations} & \\\hline
Only estimate of AI exposure comes from Eloundou et al. (2024)  & +/- \\
Limited estimates and granularity for cost savings & +/- \\
Limited granularity of energy and environmental data & +/- \\
Energy and environmental data only through 2014 & + \\
Only consider impact in the US & +/- \\
\end{tabular}
\end{table}

By assuming that AI only affects productivity by performing tasks previously performed by labor at a lower cost, we do not account for the possibility that AI could introduce new tasks or that AI could affect other forms of production, such as capital. By constraining the channels through which AI impacts productivity, we likely underestimate the aggregate impact of AI on productivity and, consequently, on energy use and carbon emissions. 

Additionally, we focus on changes in energy use driven by productivity gains from AI adoption, ignoring the possibility of AI-driven improvements in energy efficiency. If AI were to spur improvements in energy efficiency, this would likely diminish the net impact of AI on energy use and carbon emissions, implying that our estimates would be overestimates. 

Relatedly, our model assumes that prices and factors of production are fixed, so there are no changes in the composition of the economy. Thus, energy and emissions intensities are constant and there is no change in economic composition across industries. It is unclear a priori the sign of bias from these assumptions. It will likely depend on the relative energy and emissions intensities of growing and shrinking sectors, and AI adoption could alter these factors over time.

Data availability also limits our analysis. While we consider alternative AI exposure measures, they all derive from a single dataset of AI exposure estimates \cite{eloundou_gpts_2024}. Considering a broader range of exposure estimates could enhance the robustness of our findings, which could be done by applying our model to new estimates as they become available. 

Another key data limitation is the varying granularity of data. Ideally, we would have all relevant data at the task level. However, while data on AI exposure is available at the task level, energy and carbon data are at the industry-level and we apply a single uniform Cost Savings factor. Finer granularity will improve the accuracy of our estimates but it is not obvious the sign of any bias from aggregation.

For our analysis, we measure energy intensity and carbon intensity using data for 2014, the most recent available data. Yet, there has been a trend towards improved energy efficiency and decarbonization of energy systems in the US. By using older data we likely overestimate the impact of AI on energy use and carbon emissions. In the Supplementary Materials \ref{app: Projecting WIOD}, we project output, energy use, and emissions through 2023 and re-estimate the impact of AI. We find changes in energy use and carbon emissions decrease slightly to 24PJ and 790ktCO$_2$, respectively. 

While our analysis focuses on the US, similar patterns might be expected in other developed economies. A global analysis could provide a more comprehensive picture of AI's impact on energy use and emissions.

Through our parsimonious model, we capture both the direct and indirect effects of AI adoption across the economy. This provides a more comprehensive view of AI's potential impact on energy use and emissions than approaches focusing solely on the energy consumption of AI hardware or specific AI applications. Our findings indicate that, while AI adoption does increase energy use and emissions, the magnitude of this increase is relatively modest compared to overall economic activity. However, the cumulative effect over time and across sectors underscores the importance of considering energy and environmental impacts in AI development and deployment strategies. Moreover, the variation in impacts across industries highlights the need for sector-specific approaches to managing the energy and environmental consequences of AI adoption.

As AI continues to transform various sectors of the economy, we must balance productivity gains and economic benefits with potential increases in energy demand and associated emissions. This may involve prioritizing energy-efficient AI technologies, investing in renewable energy sources to power AI infrastructure, and developing strategies to offset increased emissions in AI-intensive industries.

Future research could address some of the limitations identified here, incorporating dynamic effects, exploring sector-specific AI impacts, and investigating the interplay between AI-driven productivity gains and energy efficiency improvements. Such work would further refine our understanding of AI's role in shaping future energy demand and environmental outcomes. Additionally, as more data becomes available on the real-world impacts of AI adoption across different industries, researchers can update and refine the estimates presented in this study.

It is also worth considering the potential for AI itself to contribute to solutions for energy efficiency and emissions reduction. AI technologies could play a significant role in optimizing renewable energy sources, such as wind and solar power. AI can also optimize industrial processes, increasing overall efficiency and reducing waste. Future studies should consider these aspects to provide a more comprehensive view of AI's environmental impact.

In conclusion, our study provides a valuable starting point for understanding the broader energy and environmental implications of AI adoption across the economy. We hope to stimulate further research and informed discussion on this important topic by highlighting both the potential impacts and the areas of uncertainty. As AI continues to evolve and reshape various aspects of our economy and society, ongoing analysis and monitoring of its energy and environmental impacts will ensure sustainable development of these transformative technologies.

\clearpage

\newpage

\bibliography{main.bib}

\clearpage

\appendix
\textbf{\huge Supplementary Materials}

\section{Extended Methods}\label{sec:methods}

We break our methodology down into two steps. In the first step, we estimate the exposure of industries to AI. In the second step, we merge these estimates with industry-level output, energy use, and CO$_2$ emissions data. We use these data to estimate the subsequent impact of AI adoption on GDP, energy use, and CO$_2$ emissions.

\noindent\textbf{Key Data Sources}
\begin{itemize}
    \item AI exposure: Data on exposure to AI, comes from estimates of task-level exposure developed by \citet{eloundou_gpts_2024}.
    \item Occupation by industry Wages: Data on annual wage bills at the occupation by industry level comes from the Bureau of Labor Statistics Occupational Employment and Wage Statistics \citep{us_bureau_of_labor_statistics_last_occupational_2024}.
    \item Industry-level output, energy use, and emissions: Data on output at the industry-level comes from the World Input-Output Database 2016 release \cite{wiod_world_2021,timmer_illustrated_2015}. Corresponding data on energy use and emissions at the industry-level comes from the World Input-Output Database's corresponding Environmental Accounts \cite{corsatea_world_2019}.
\end{itemize}

\noindent\textbf{Industry-level exposure to AI}

We begin with a task-level exposure developed by \citet{eloundou_gpts_2024}. In the analysis in the text we use a modified version of Eloundou et al.'s \textit{automation} measure which is a score of exposure ranging \{0, 0.25, 0.5, 0.75, 1\} where 0 is the lowest exposure and 1 is the highest exposure. Following \citet{acemoglu_simple_2024}, we convert this measure into a binary measure of whether a task is exposed. For our central estimates, we label a task as exposed if the automation score is greater than 0.5 and label all other tasks as not exposed. To capture a range of uncertainty in exposure, we also construct a lower and upper bound. For the lower bound, we label tasks as exposed only if the automation score is 1. For the upper bound, we label tasks as exposed if the automation score is greater than 0.

Taking our label of exposed tasks, we next aggregate exposure to AI to the occupational level by taking the simple average of our measure of exposed tasks across each of the tasks that compose an occupation where occupations are defined at the 8-digit SOC level. Each occupation can be composed of multiple tasks but each task matches to a specific occupation. This gives a fractional measure of the exposure of an occupation to AI based on the exposure of the occupation's underlying tasks.

To go from exposure at the occupation level to exposure at the industry-level, we match occupations to industries and take the weighted average of the exposure of occupations within an industry. For weights, we use the occupation's share of the industry's wage bill. Throughout, we try to preserve the highest level of granularity possible. Wage bills come from the Bureau of Labor Statistics, described above. Data are at the occupation by 4-digit NAICS industry classification averaged over the years 2019 through 2022. Wages are deflated from current USD to 2017 USD using the annual GDP Deflator from the US Bureau of Economic Analysis \citep{us_bureau_of_economic_analysis_gross_nodate}. To match occupations to industries, a many-to-many match, we first aggregate our measure of exposure from the 8-digit SOC level to the 6-digit SOC level, again taking the simple average. Occupations at the 6-digit SOC level are merged to industry-by-occupation wage bill data. However, for some occupations, wage data is only available at the 5-digit SOC level. Thus, we aggregate our measure of wage-bill weighted exposure to the 5-digit SOC level and merge again with the wage data for those missing data at the 6-digit level. Any remaining occupations without an exposure measure are assumed to have no exposure to AI. Finally, we add up the wage-bill weighted exposure across occupations within industries to get our measure of industry-level exposure.

The Cost Savings factor is composed of the product of the fraction of exposed tasks that are feasibly and profitably automated and the average fraction of labor cost savings from adopting AI. Labor cost savings from AI adoption (27\%) comes from the average of estimates from experimental studies in \citet{noy_experimental_2023} and \citet{brynjolfsson_generative_2023}. Lacking more granular data, we extrapolate these labor cost savings across all technologies and tasks. Applying the estimate of \citet{svanberg_beyond_2024}, we assume that 23\% of exposed tasks that are feasibly and profitably automated within 10 years. To convert this to an annual metric, we divide by 10, suggesting that 2.3\% more of exposed tasks are feasibly and profitably automated each year. This estimate comes from the setting of computer vision technologies, but given that this is the only such estimate to our knowledge, we again extrapolate this to apply across all AI technologies and tasks. We explore the sensitivity of our estimates to these key parameters in Figure \ref{fig:sensitivity}.

\noindent\textbf{Industry-level impacts of AI adoption}

We convert our estimates of industry-level productivity gains from AI adoption into broader impacts on GDP, energy use, and emissions we gather industry-level data on output (value added), energy intensity, and emissions intensity from the World Input-Output Database and corresponding Environmental Accounts. We convert the industry-level output data from current USD to 2017USD using the US GDP deflator from the US Bureau of Economic Analysis \citep{us_bureau_of_economic_analysis_gross_nodate}.

The World Input-Output Database and corresponding Environmental Accounts provide information on output (value added), energy use, and CO$_2$ emissions across 56 ISIC rev. 4 industry classifications. Of these 56 industries, we use 55 industries, dropping industry code U: ``Activities of extraterritorial organizations and bodies''. To connect our measures of exposure to AI at the 4-digit NAICS industry classification to the ISIC rev. 4 industry classifications of the World Input-Output database, we use the concord\_naics\_isic(.) function from the concordance R package version 2.1.0 written by Liao et al. (2020). We first match 4-digit NAICS industry classifications to 2-digit ISIC rev. 4 industry classifications using the match of the largest share of occurrences, a many-to-one match. We then aggregate across the wage-bill weighted exposure of 4-digit NAICS industry classifications within each 2-digit ISIC rev. 4 industry classification and again to match the industry classifications of the World Input-Output Database which can be an aggregate of multiple 2-digit ISIC rev. 4 industries.

To estimate the industry-level and aggregate change in energy use due to the productivity gains from AI adoption, we multiply the change in industry-level output by each industry's energy intensity. To calculate the change in emissions, we further multiply by the emissions intensity of each industry. Energy intensity of an industry is measured as the the energy use of that industry divided by its output. Emissions intensity of an industry is measured as the emissions of that industry divided by its energy use. For both energy intensity and emissions intensity, we use data for 2014, the latest year in the WIOD data. Given that both energy intensity and emissions intensity are declining over time, in the Supplementary Materials \ref{app: Projecting WIOD} we present estimates using projected energy intensity and emissions intensity for the year 2023. These are constructed by linearly projecting output, energy use, and emissions for each industry.

\section{Baseline Estimates}\label{app: Baseline Estimates}

\footnotesize
\begin{longtable}{lcccc}
\caption{Changes in GDP, Energy, and Emissions by Industry} \label{tab:industry_changes} \\
\toprule
Industry & Code & Change in GDP & Change in Energy & Change in Emissions \\
\midrule
\endfirsthead

\multicolumn{5}{c}%
{{\bfseries \tablename\ \thetable{} -- continued from previous page}} \\
\toprule
Industry & Code & Change in GDP & Change in Energy & Change in Emissions \\
\midrule
\endhead

\midrule \multicolumn{5}{r}{{Continued on next page}} \\ \midrule
\endfoot

\bottomrule
\endlastfoot
Fishing and aquaculture & A03 & 0 & 0 & 0 \\
 & & [0, 0] & [0, 0] & [0, 0] \\
\midrule
Public administration and defence & O84 & 0 & 0 & 0 \\
 & & [0, 0] & [0, 0] & [0, 0] \\
\midrule
Activities of households as employers & T & 0 & 0 & 0 \\
 & & [0, 0] & [0, 0] & [0, 0] \\
\midrule
Manufacture of coke and refined petroleum & C19 & 0.00940 & 0.45008 & 1.48014 \\
 & & [0.00015, 0.03526] & [0.00741, 1.68908] & [0.02436, 5.55475] \\
\midrule
Crop and animal production & A01 & 0.00958 & 0.01830 & 1.12724 \\
 & & [0.00013, 0.03208] & [0.00025, 0.06126] & [0.01535, 3.77319] \\
\midrule
Mining and quarrying & B & 0.03696 & 0.13969 & 6.99116 \\
 & & [0.00053, 0.13886] & [0.00201, 0.52487] & [0.10043, 26.26915] \\
\midrule
Manufacture of motor vehicles & C29 & 0.03565 & 0.01203 & 0.31706 \\
 & & [0.00049, 0.12859] & [0.00017, 0.04339] & [0.00439, 1.14355] \\
\midrule
Real estate activities & L68 & 0.17219 & 0.02801 & 0.77174 \\
 & & [0.00219, 0.43398] & [0.00036, 0.07060] & [0.00981, 1.94505] \\
\midrule
Manufacture of basic metals & C24 & 0.01794 & 0.10980 & 6.97384 \\
 & & [0.00021, 0.06019] & [0.00131, 0.36831] & [0.08292, 23.39344] \\
\midrule
Manufacture of chemicals & C20 & 0.03983 & 0.38278 & 10.32334 \\
 & & [0.00077, 0.14094] & [0.00743, 1.35446] & [0.20033, 36.52883] \\
\midrule
Forestry and logging & A02 & 0.00245 & 0.00435 & 0.19736 \\
 & & [0.00002, 0.00512] & [0.00003, 0.00911] & [0.00149, 0.41330] \\
\midrule
Manufacture of food products & C10\_12 & 0.07921 & 0.10458 & 4.65548 \\
 & & [0.00213, 0.23178] & [0.00281, 0.30602] & [0.12528, 13.62278] \\
\midrule
Water transport & H50 & 0.00574 & 0.08202 & 5.63490 \\
 & & [0.00015, 0.01964] & [0.00213, 0.28077] & [0.14601, 19.28956] \\
\midrule
Manufacture of paper and paper products & C17 & 0.01869 & 0.18512 & 3.01534 \\
 & & [0.00049, 0.05733] & [0.00488, 0.56780] & [0.07955, 9.24873] \\
\midrule
Manufacture of basic pharmaceutical products & C21 & 0.02860 & 0.01426 & 0.12963 \\
 & & [0.00058, 0.10696] & [0.00029, 0.05331] & [0.00265, 0.48472] \\
\midrule
Electricity, gas, steam and air conditioning supply & D35 & 0.05458 & 5.13249 & 272.83147 \\
 & & [0.00065, 0.18354] & [0.06128, 17.26050] & [3.25736, 917.52820] \\
\midrule
Manufacture of wood and wood products & C16 & 0.01416 & 0.04015 & 0.82426 \\
 & & [0.00018, 0.04089] & [0.00051, 0.11596] & [0.01041, 2.38056] \\
\midrule
Manufacture of rubber and plastic products & C22 & 0.03460 & 0.12468 & 4.34824 \\
 & & [0.00069, 0.10825] & [0.00248, 0.39005] & [0.08641, 13.60302] \\
\midrule
Telecommunications & J61 & 0.09917 & 0.01104 & 0.54065 \\
 & & [0.00265, 0.24268] & [0.00029, 0.02702] & [0.01442, 1.32305] \\
\midrule
Manufacture of textiles & C13\_15 & 0.01486 & 0.02220 & 0.68690 \\
 & & [0.00027, 0.04085] & [0.00041, 0.06104] & [0.01266, 1.88851] \\
\midrule
Manufacture of other transport equipment & C30 & 0.05793 & 0.04294 & 1.64789 \\
 & & [0.00067, 0.22011] & [0.00050, 0.16313] & [0.01908, 6.26122] \\
\midrule
Manufacture of machinery and equipment n.e.c. & C28 & 0.07340 & 0.10064 & 4.04501 \\
 & & [0.00076, 0.24358] & [0.00104, 0.33400] & [0.04168, 13.42378] \\
\midrule
Manufacture of other non-metallic mineral products & C23 & 0.02179 & 0.14473 & 20.72346 \\
 & & [0.00041, 0.06182] & [0.00274, 0.41061] & [0.39290, 58.79407] \\
\midrule
Manufacture of electrical equipment & C27 & 0.02498 & 0.02335 & 0.78101 \\
 & & [0.00031, 0.08270] & [0.00029, 0.07731] & [0.00962, 2.58539] \\
\midrule
Manufacture of fabricated metal products & C25 & 0.07813 & 0.08393 & 1.93780 \\
 & & [0.00103, 0.24034] & [0.00110, 0.25818] & [0.02543, 5.96108] \\
\midrule
Insurance, reinsurance and pension funding & K65 & 0.24235 & 0.03763 & 2.39246 \\
 & & [0.00269, 0.54819] & [0.00042, 0.08512] & [0.02658, 5.41171] \\
\midrule
Manufacture of furniture & C31\_32 & 0.06411 & 0.07142 & 2.37054 \\
 & & [0.00092, 0.18817] & [0.00102, 0.20963] & [0.03391, 6.95753] \\
\midrule
Motion picture, video and television programme production & J59\_60 & 0.09006 & 0.00166 & 0.04684 \\
 & & [0.00179, 0.23383] & [0.00003, 0.00430] & [0.00093, 0.12163] \\
\midrule
Land transport and transport via pipelines & H49 & 0.15201 & 1.09109 & 70.79556 \\
 & & [0.00312, 0.33645] & [0.02240, 2.41503] & [1.45369, 156.69877] \\
\midrule
Construction & F & 0.36589 & 0.65634 & 18.44284 \\
 & & [0.00901, 1.37906] & [0.01616, 2.47380] & [0.45404, 69.51271] \\
\midrule
Water collection, treatment and supply & E36 & 0.00426 & 0.10949 & 5.52249 \\
 & & [0.00008, 0.01238] & [0.00201, 0.31814] & [0.10117, 16.04610] \\
\midrule
Sewerage; waste collection, treatment and disposal activities & E37\_39 & 0.02895 & 0.05231 & 2.39245 \\
 & & [0.00051, 0.07335] & [0.00092, 0.13253] & [0.04213, 6.06164] \\
\midrule
Advertising and market research & M73 & 0.07745 & 0.03641 & 2.14277 \\
 & & [0.00123, 0.20412] & [0.00058, 0.09595] & [0.03401, 5.64712] \\
\midrule
Manufacture of computer, electronic and optical products & C26 & 0.12672 & 0.04183 & 1.54658 \\
 & & [0.00118, 0.43769] & [0.00039, 0.14450] & [0.01442, 5.34211] \\
\midrule
Wholesale trade, except of motor vehicles and motorcycles & G46 & 0.53526 & 0.14432 & 8.80161 \\
 & & [0.00914, 1.36541] & [0.00246, 0.36815] & [0.15027, 22.45240] \\
\midrule
Other service activities & R\_S & 0.29747 & 0.20467 & 12.27641 \\
 & & [0.00751, 0.86650] & [0.00517, 0.59619] & [0.30996, 35.75997] \\
\midrule
Architectural and engineering activities & M71 & 0.18470 & 0.08682 & 5.10978 \\
 & & [0.00202, 0.71120] & [0.00095, 0.33431] & [0.05584, 19.67546] \\
\midrule
Printing and reproduction of recorded media & C18 & 0.03560 & 0.03836 & 1.18287 \\
 & & [0.00281, 0.07818] & [0.00302, 0.08425] & [0.09327, 2.59795] \\
\midrule
Air transport & H51 & 0.09041 & 1.31683 & 89.96491 \\
 & & [0.00908, 0.17330] & [0.13228, 2.52423] & [9.03702, 172.45441] \\
\midrule
Accommodation and food service activities & I & 0.42864 & 0.46834 & 28.28131 \\
 & & [0.02984, 1.09179] & [0.03260, 1.19292] & [1.96859, 72.03582] \\
\midrule
Scientific research and development & M72 & 0.12593 & 0.05920 & 3.48401 \\
 & & [0.00140, 0.47704] & [0.00066, 0.22424] & [0.03873, 13.19745] \\
\midrule
Human health and social work activities & Q & 1.10139 & 0.69661 & 42.94767 \\
 & & [0.01981, 5.07354] & [0.01253, 3.20889] & [0.77242, 197.83739] \\
\midrule
Administrative and support service activities & N & 0.63764 & 0.54030 & 33.77095 \\
 & & [0.01914, 1.53223] & [0.01622, 1.29832] & [1.01368, 81.15048] \\
\midrule
Repair and installation of machinery and equipment & C33 & 0.02357 & 0.03884 & 1.71706 \\
 & & [0.00046, 0.06743] & [0.00076, 0.11114] & [0.03343, 4.91306] \\
\midrule
Legal and accounting activities & M69\_70 & 0.75134 & 0.22207 & 12.93869 \\
 & & [0.01294, 1.82427] & [0.00382, 0.53919] & [0.22277, 31.41552] \\
\midrule
Warehousing and support activities for transportation & H52 & 0.13151 & 0.21378 & 13.73676 \\
 & & [0.00459, 0.31997] & [0.00746, 0.52016] & [0.47950, 33.42312] \\
\midrule
Wholesale and retail trade and repair of motor vehicles & G45 & 0.29578 & 0.16105 & 9.71064 \\
 & & [0.00488, 0.76903] & [0.00266, 0.41874] & [0.16023, 25.24786] \\
\midrule
Computer programming, consultancy and related activities & J62\_63 & 0.59254 & 0.23704 & 14.64565 \\
 & & [0.00740, 1.57959] & [0.00296, 0.63190] & [0.18281, 39.04209] \\
\midrule
Activities auxiliary to financial services and insurance & K66 & 0.54175 & 0.07790 & 4.77260 \\
 & & [0.00391, 1.14307] & [0.00056, 0.16437] & [0.03448, 10.06997] \\
\midrule
Other professional, scientific and technical activities & M74\_75 & 0.10090 & 0.04743 & 2.79147 \\
 & & [0.00230, 0.27584] & [0.00108, 0.12966] & [0.06352, 7.63116] \\
\midrule
Postal and courier activities & H53 & 0.14106 & 0.15354 & 8.76502 \\
 & & [0.00615, 0.27075] & [0.00669, 0.29471] & [0.38215, 16.82353] \\
\midrule
Retail trade, except of motor vehicles and motorcycles & G47 & 1.72690 & 1.27943 & 78.77955 \\
 & & [0.19333, 3.49253] & [0.14324, 2.58757] & [8.81953, 159.32642] \\
\midrule
Financial service activities, except insurance & K64 & 1.02793 & 0.25678 & 16.18083 \\
 & & [0.02414, 2.32556] & [0.00603, 0.58094] & [0.38004, 36.60699] \\
\midrule
Publishing activities & J58 & 0.53142 & 0.00274 & 0.07993 \\
 & & [0.00726, 1.36713] & [0.00004, 0.00705] & [0.00109, 0.20564] \\
\midrule
Education & P85 & 0.77359 & 12.47747 & 51.13327 \\
 & & [0.01298, 3.61802] & [0.20943, 58.35591] & [0.85826, 239.14526] \\
\end{longtable}
\normalsize

\newpage

\section{Projecting WIOD}\label{app: Projecting WIOD}
For data on energy use and emissions at the industry level, we use the Environmental Accounts from the World Input-Output Database 2016 release's corresponding Environmental Accounts. This data covers the years 2000 to 2016. For our analysis in the text, we use data for the year 2014 since it is the most recent year of economic data available. However, using older data potentially introduces bias in our analysis (Table 2). Specifically, energy intensity and emissions intensity change over time, on the aggregate level generally declining with time (Figure 5ab). This suggests that using more recent data on energy and emissions would lead to lower estimates for the energy and emissions impact of changes in productivity due to AI adoption. 

To get at the potential scale of this bias, we use a simple projection of industry-level output, energy use, and emissions. For each industry and variable, we fit a log-linear curve to the data from 2000 to 2014 and use this fit to project the variable out to 2023 (Figure 4). We then use the projected values to recalculate industry-level energy intensity and emissions intensity (Figure 5). Finally, we use the projected values for 2023 to recalculate the impact of AI adoption on energy use and CO$_2$ emissions.

Using the new projected values for industry-level energy intensity and emissions intensity we recalculate the aggregate change in energy and aggregate change in emissions corresponding with estimates in Equation (3) in the text. We find, 

\begin{equation*}
    \Delta E=\sum_{k=1}^N \Delta E_k=24 PJ\text{, and } \Delta C=\sum_{k=1}^N \Delta C_k = 790 ktCO_2
\end{equation*}

As expected, these estimates of change in energy use and change in emissions due to AI adoption are lower than the estimates in the text which use older, higher values for energy and emissions intensity. Table 4 illustrates this for the same select industries shown in Table 1. However, the differences in the estimates are arguably small compared with the broader uncertainties in our modeling approach.

\begin{figure}[h!]
    \centering
    \begin{subfigure}{0.3\textwidth}
    \centering
    \includegraphics[width=0.5\linewidth]{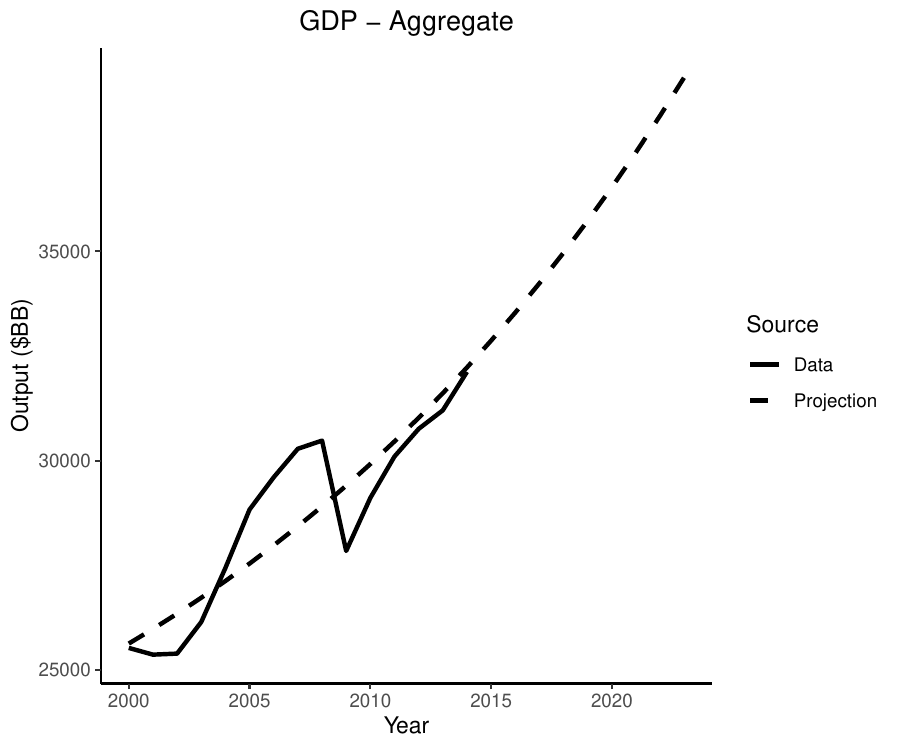}
    \caption{}
    \end{subfigure}%
    \begin{subfigure}{0.3\textwidth}
    \centering
    \includegraphics[width=0.5\linewidth]{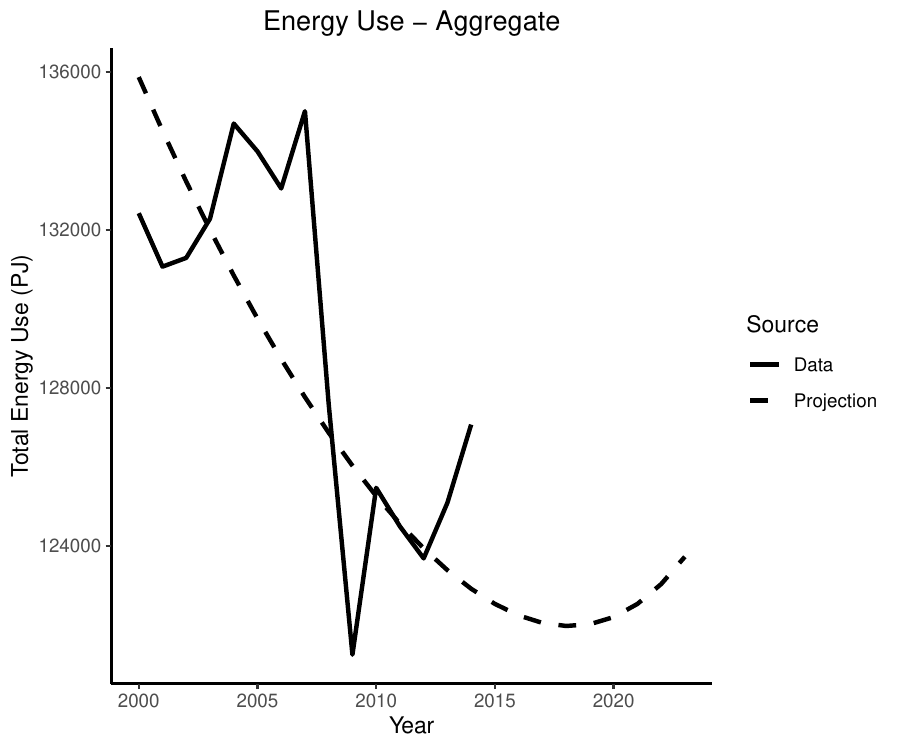}
    \caption{}
    \end{subfigure}%
    \begin{subfigure}{0.3\textwidth}
    \centering
    \includegraphics[width=0.5\linewidth]{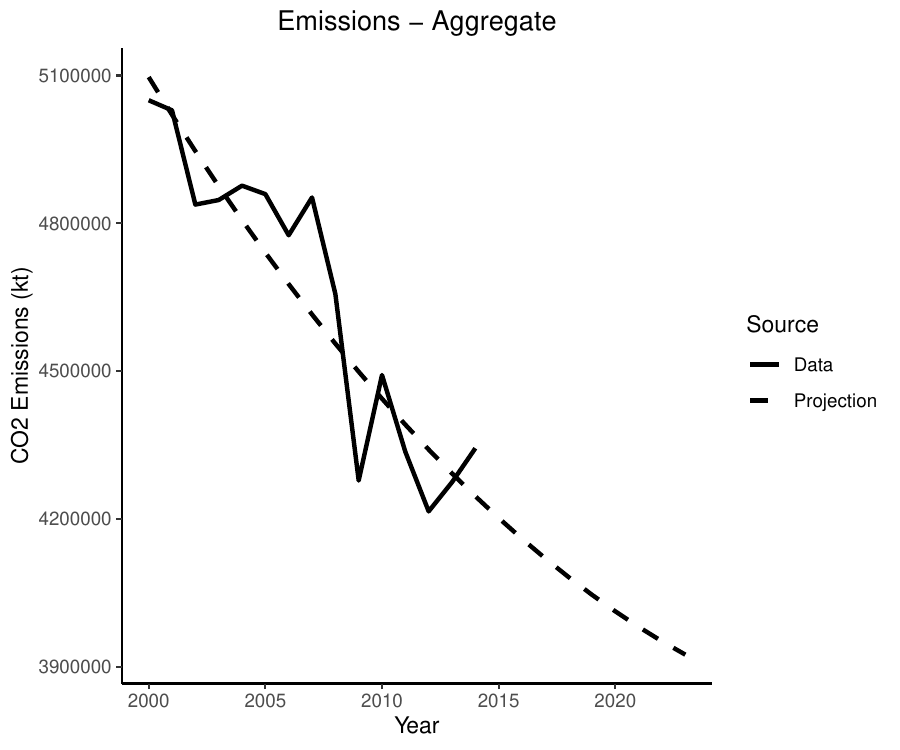}
    \caption{}
    \end{subfigure}
    
    \begin{subfigure}{0.3\textwidth}
    \centering
    \includegraphics[width=0.5\linewidth]{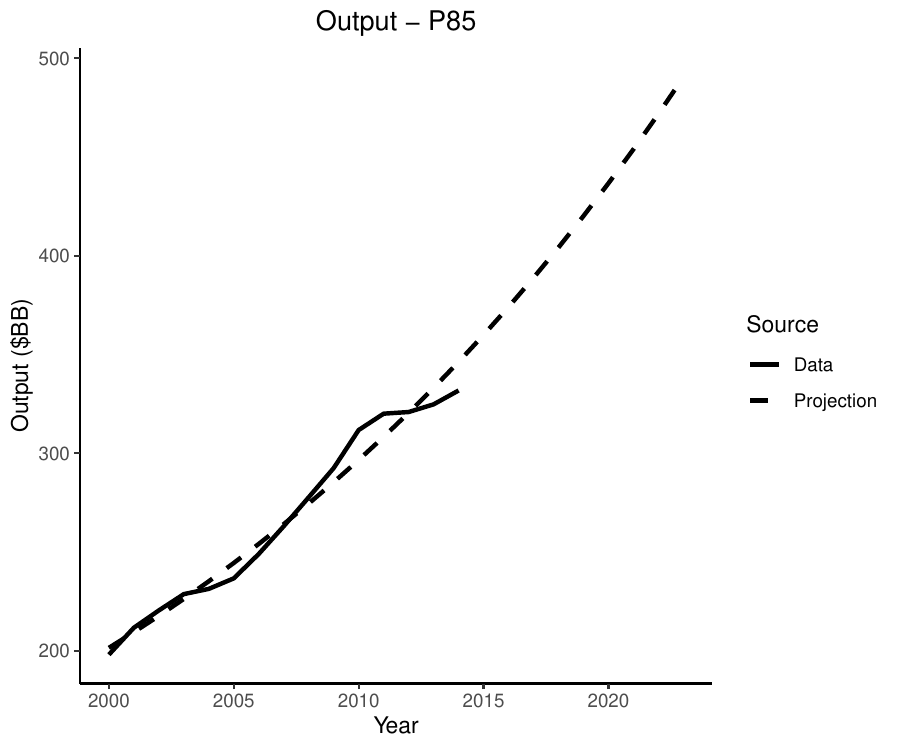}
    \caption{}
    \end{subfigure}%
    \begin{subfigure}{0.3\textwidth}
    \centering
    \includegraphics[width=0.5\linewidth]{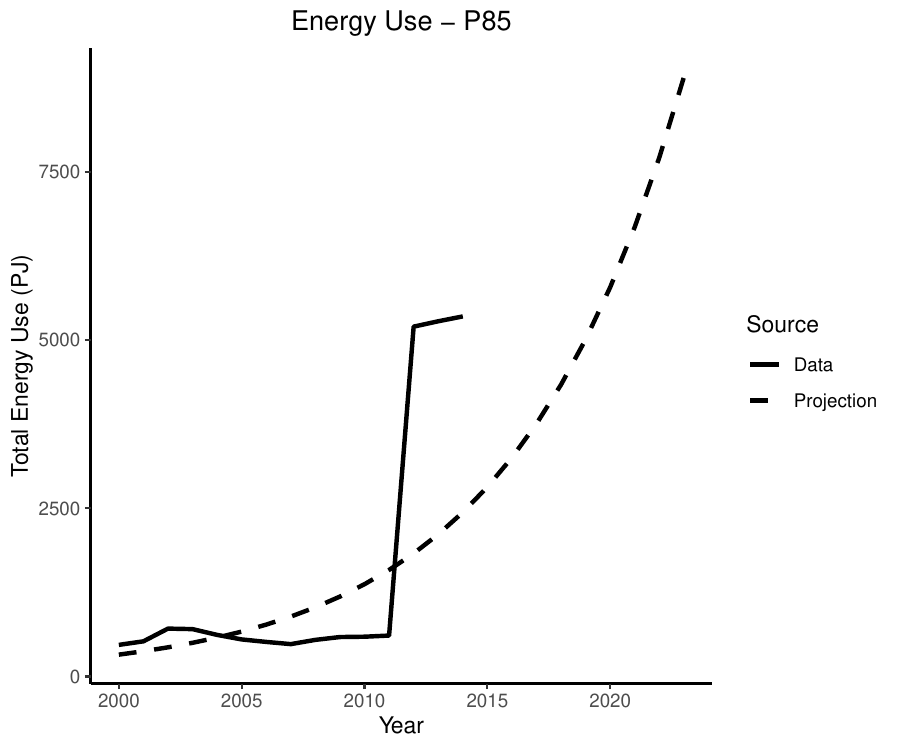}
    \caption{}
    \end{subfigure}%
    \begin{subfigure}{0.3\textwidth}
    \centering
    \includegraphics[width=0.5\linewidth]{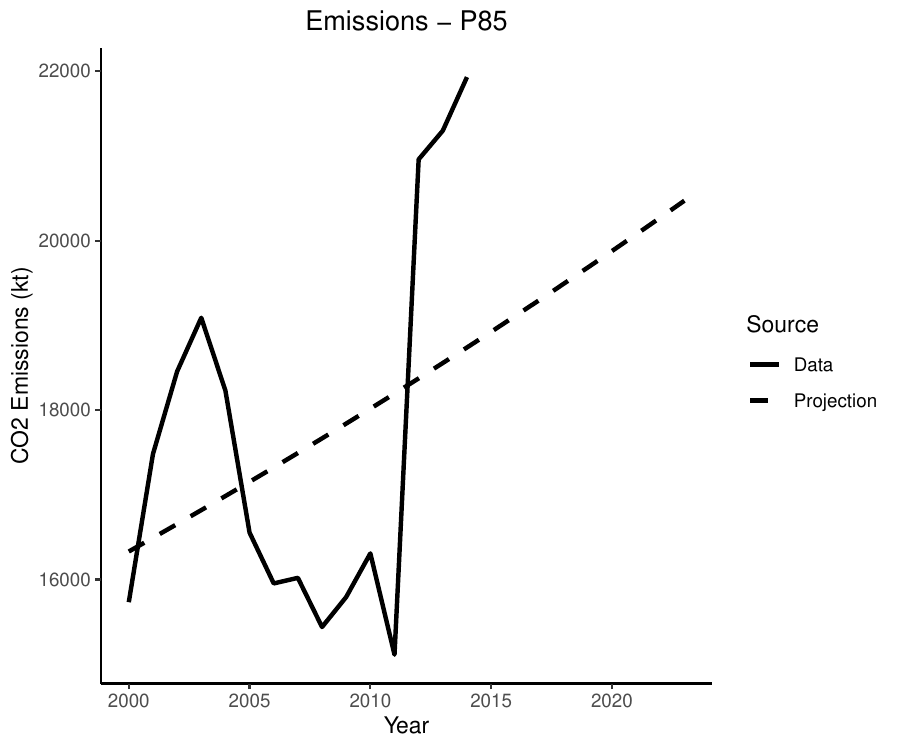}
    \caption{}
    \end{subfigure}

    \begin{subfigure}{0.3\textwidth}
    \centering
    \includegraphics[width=0.5\linewidth]{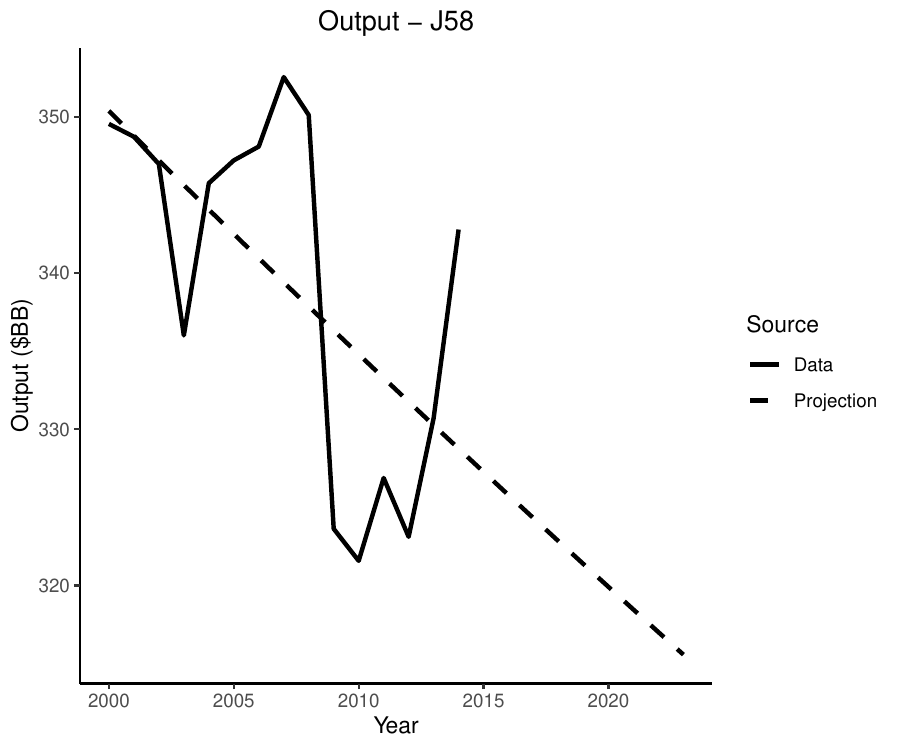}
    \caption{}
    \end{subfigure}%
    \begin{subfigure}{0.3\textwidth}
    \centering
    \includegraphics[width=0.5\linewidth]{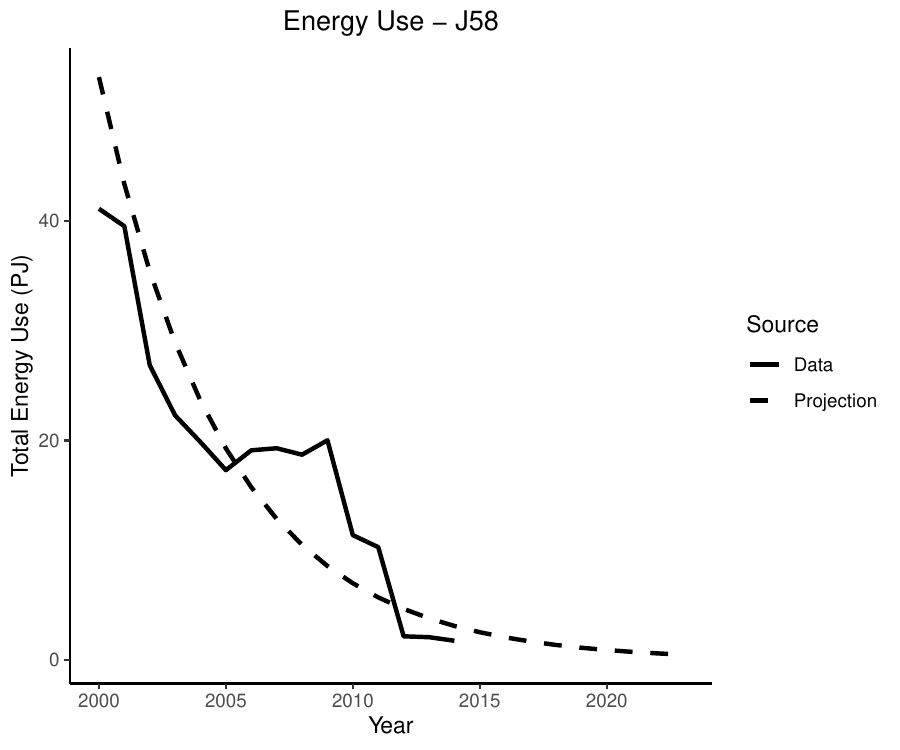}
    \caption{}
    \end{subfigure}%
    \begin{subfigure}{0.3\textwidth}
    \centering
    \includegraphics[width=0.5\linewidth]{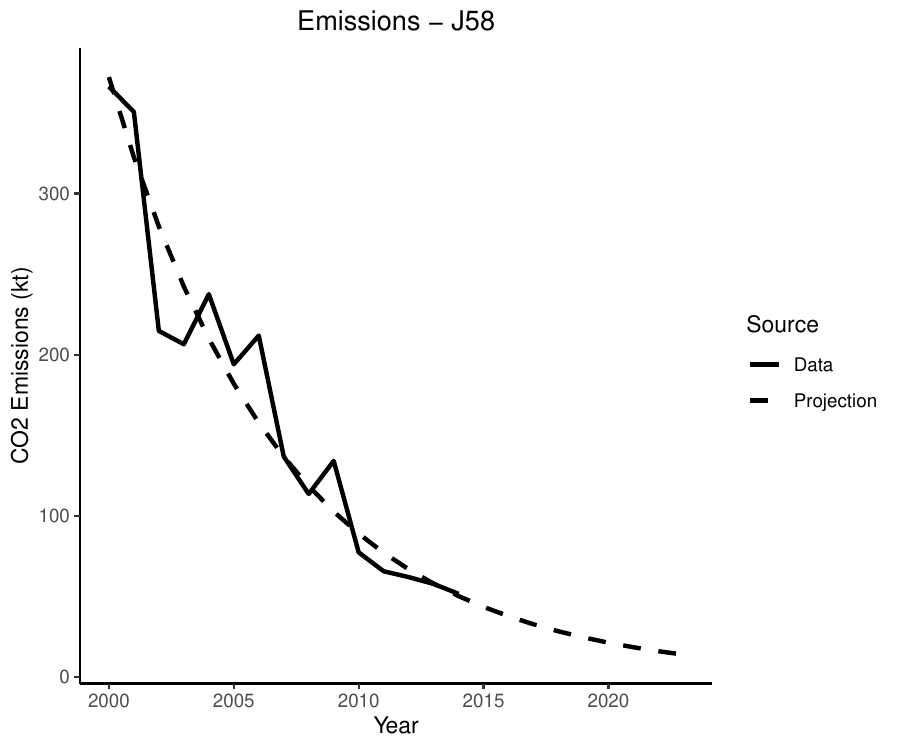}
    \caption{}
    \end{subfigure}

    \begin{subfigure}{0.3\textwidth}
    \centering
    \includegraphics[width=0.5\linewidth]{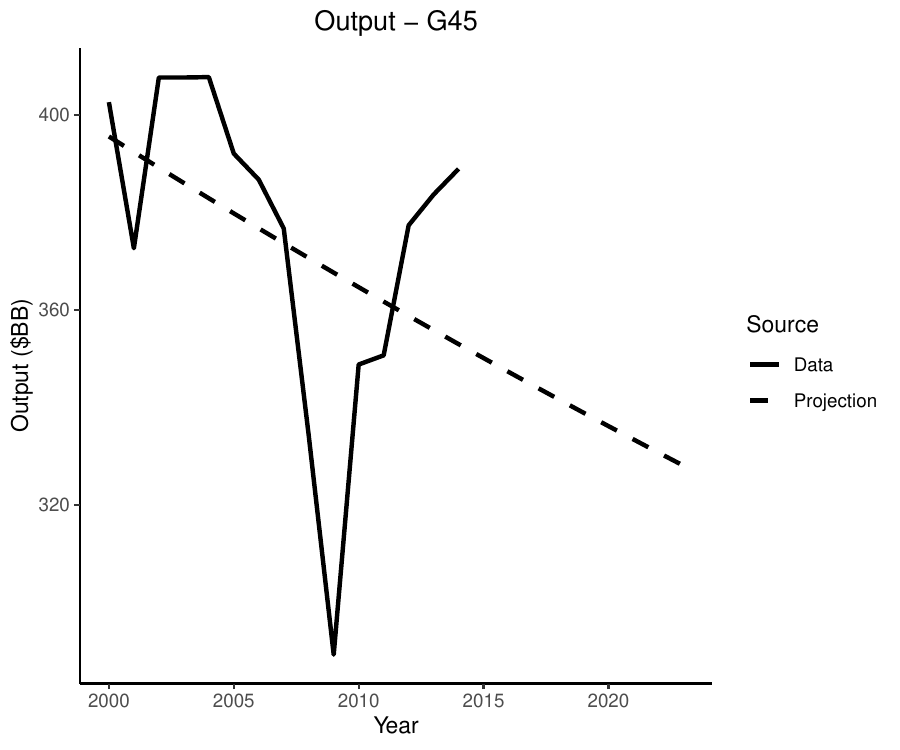}
    \caption{}
    \end{subfigure}%
    \begin{subfigure}{0.3\textwidth}
    \centering
    \includegraphics[width=0.5\linewidth]{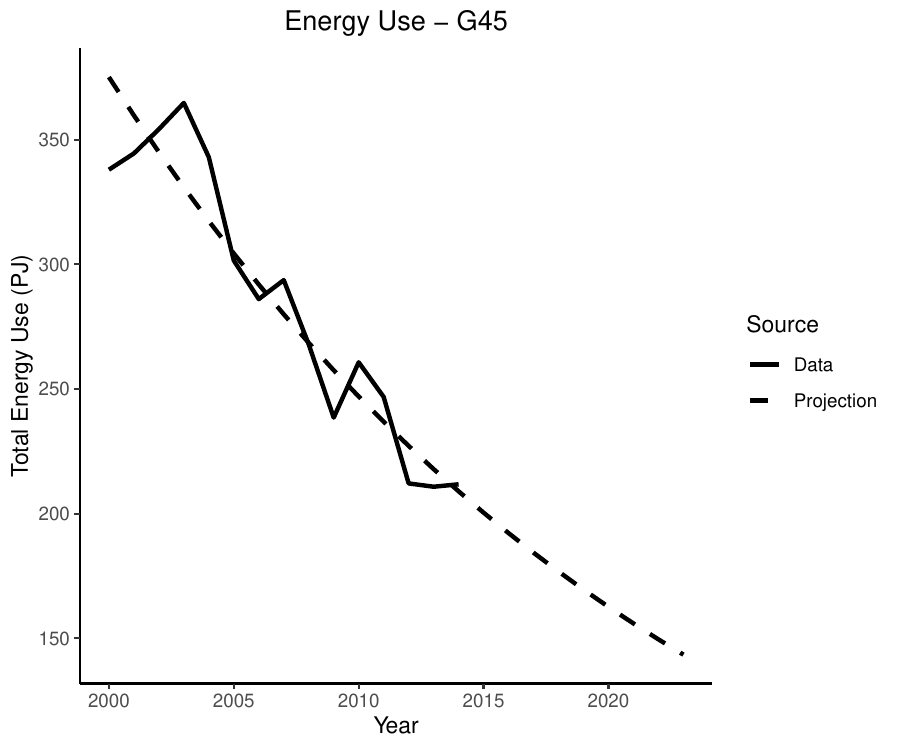}
    \caption{}
    \end{subfigure}%
    \begin{subfigure}{0.3\textwidth}
    \centering
    \includegraphics[width=0.5\linewidth]{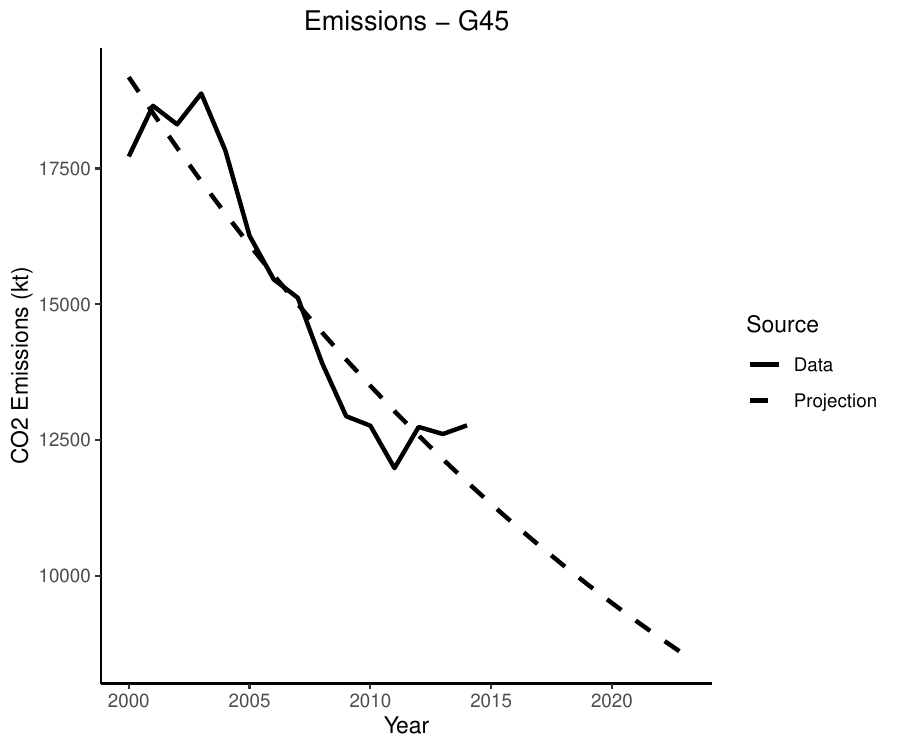}
    \caption{}
    \end{subfigure}
    
    \caption{Projection of Output, Energy Use, and Emissions}
    \label{fig:WIOD Projections}
\end{figure}

\begin{figure}[h!]
    \centering
    
    \begin{subfigure}{0.5\textwidth}
    \centering
    \includegraphics[width=0.5\linewidth]{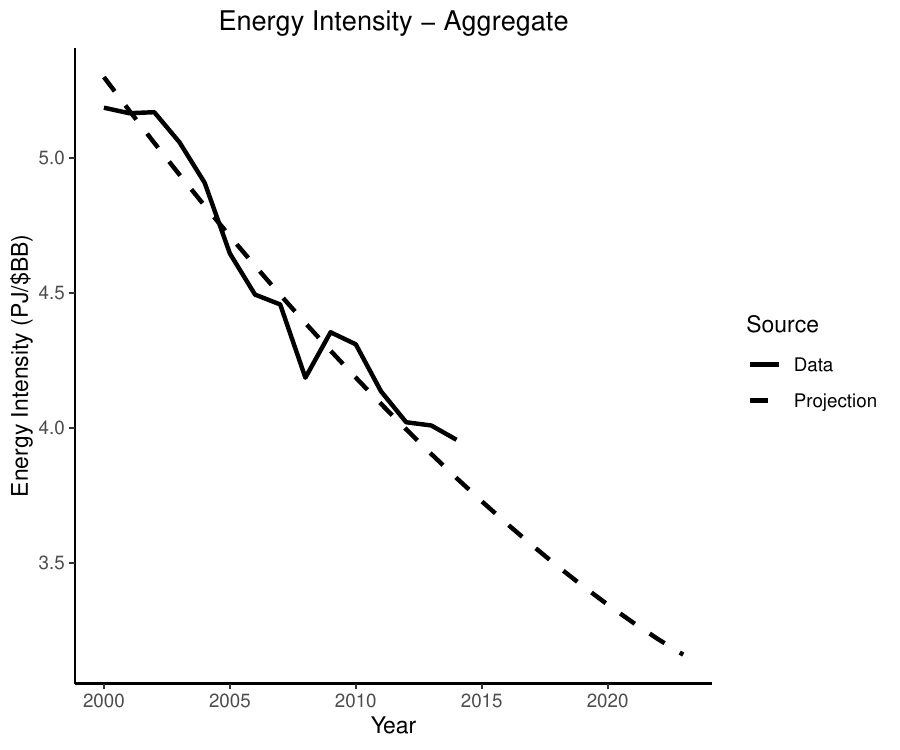}
    \caption{}
    \end{subfigure}%
    \begin{subfigure}{0.5\textwidth}
    \centering
    \includegraphics[width=0.5\linewidth]{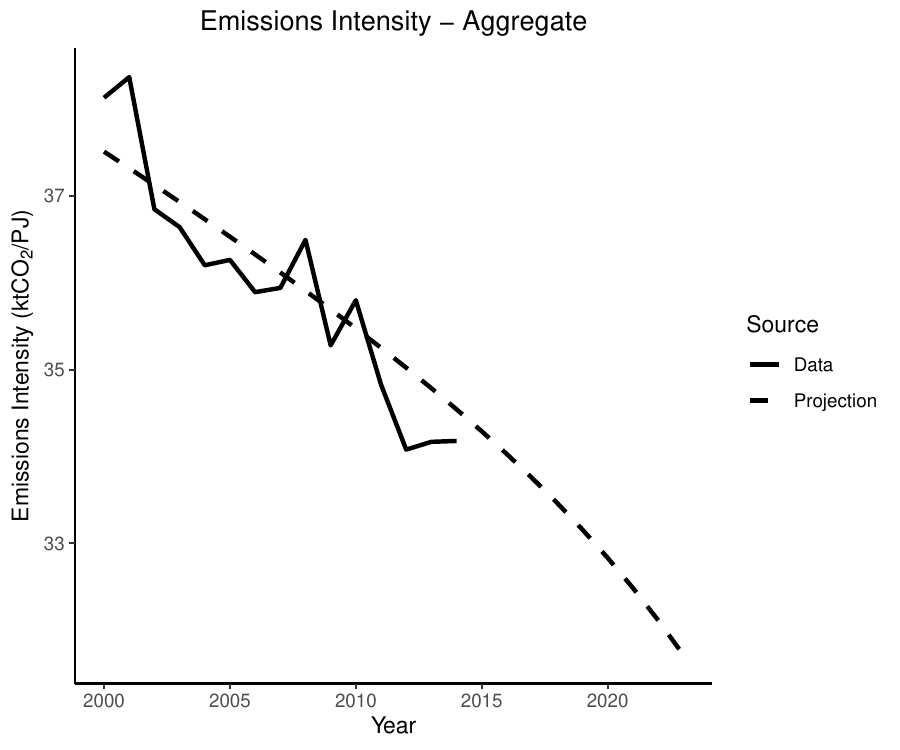}
    \caption{}
    \end{subfigure}
    
    \begin{subfigure}{0.5\textwidth}
    \centering
    \includegraphics[width=0.5\linewidth]{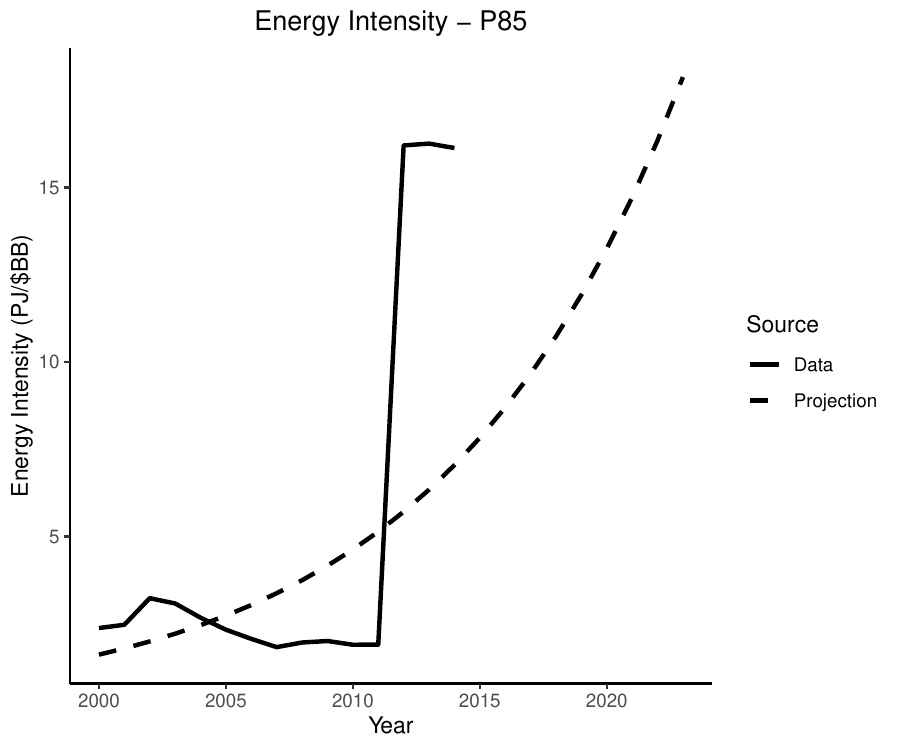}
    \caption{}
    \end{subfigure}%
    \begin{subfigure}{0.5\textwidth}
    \centering
    \includegraphics[width=0.5\linewidth]{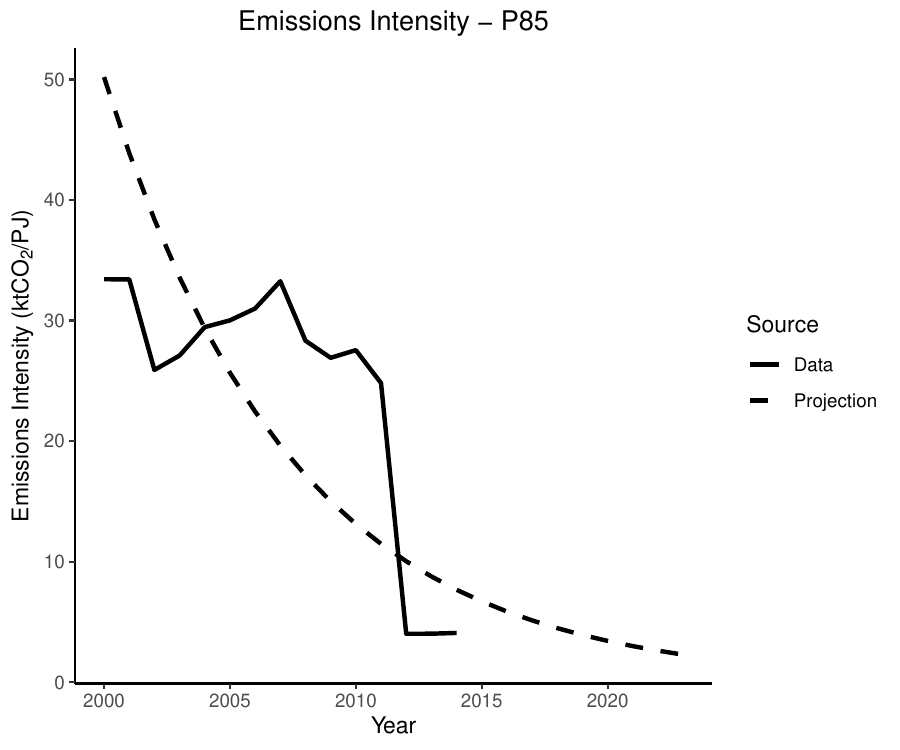}
    \caption{}
    \end{subfigure}
    
    \begin{subfigure}{0.5\textwidth}
    \centering
    \includegraphics[width=0.5\linewidth]{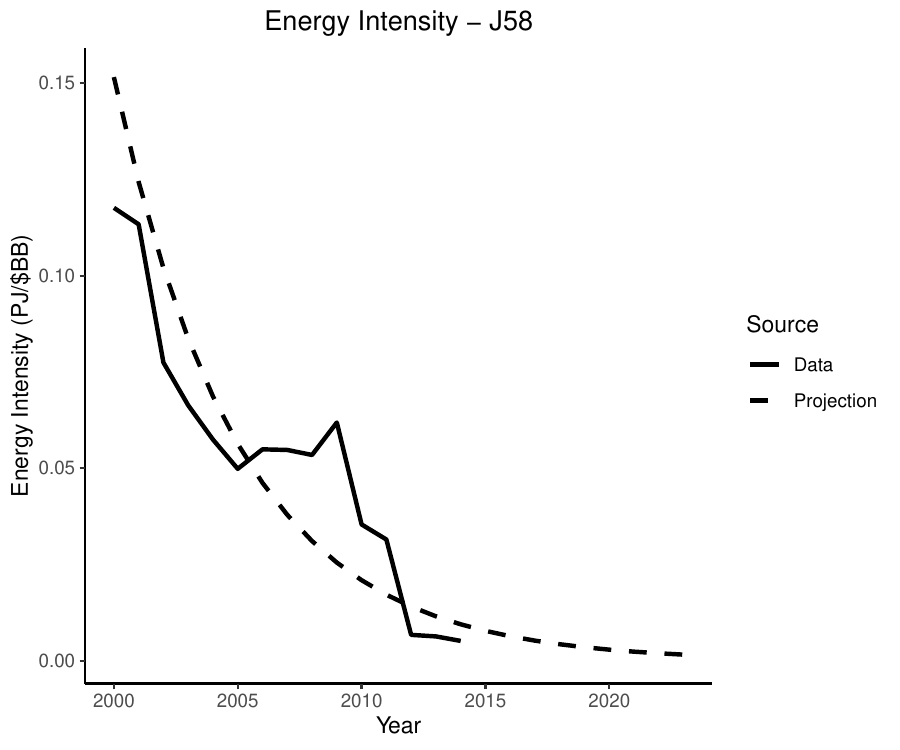}
    \caption{}
    \end{subfigure}%
    \begin{subfigure}{0.5\textwidth}
    \centering
    \includegraphics[width=0.5\linewidth]{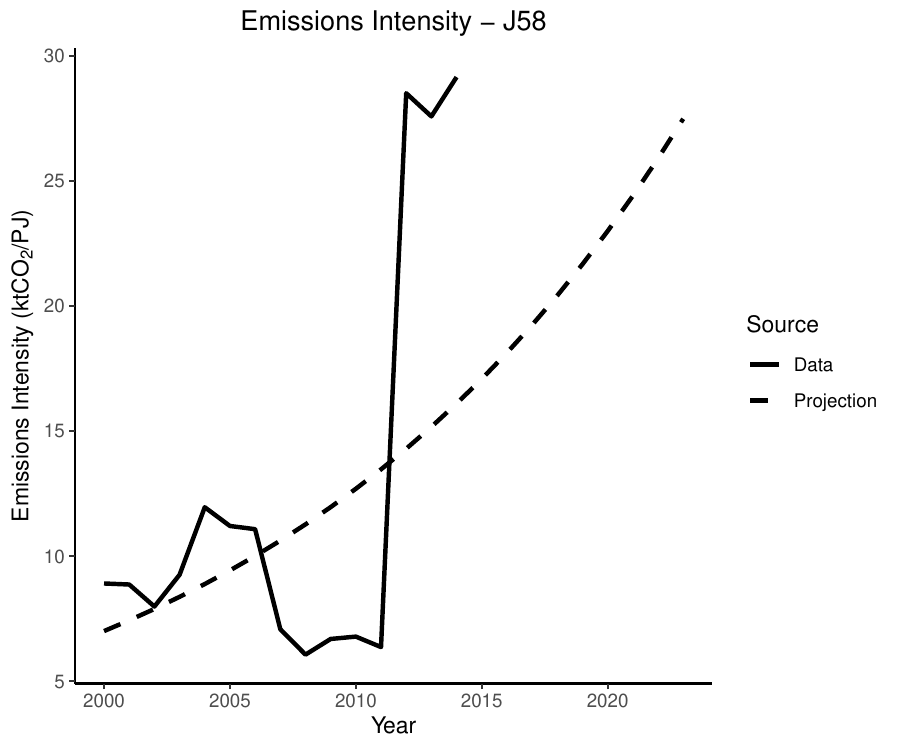}
    \caption{}
    \end{subfigure}
    
    \begin{subfigure}{0.5\textwidth}
    \centering
    \includegraphics[width=0.5\linewidth]{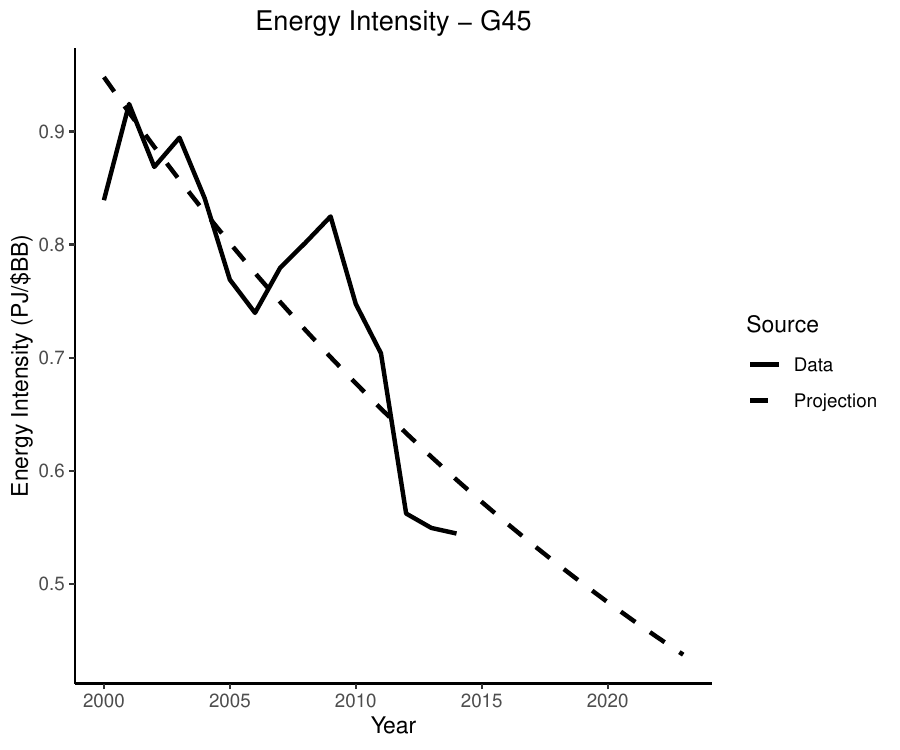}
    \caption{}
    \end{subfigure}%
    \begin{subfigure}{0.5\textwidth}
    \centering
    \includegraphics[width=0.5\linewidth]{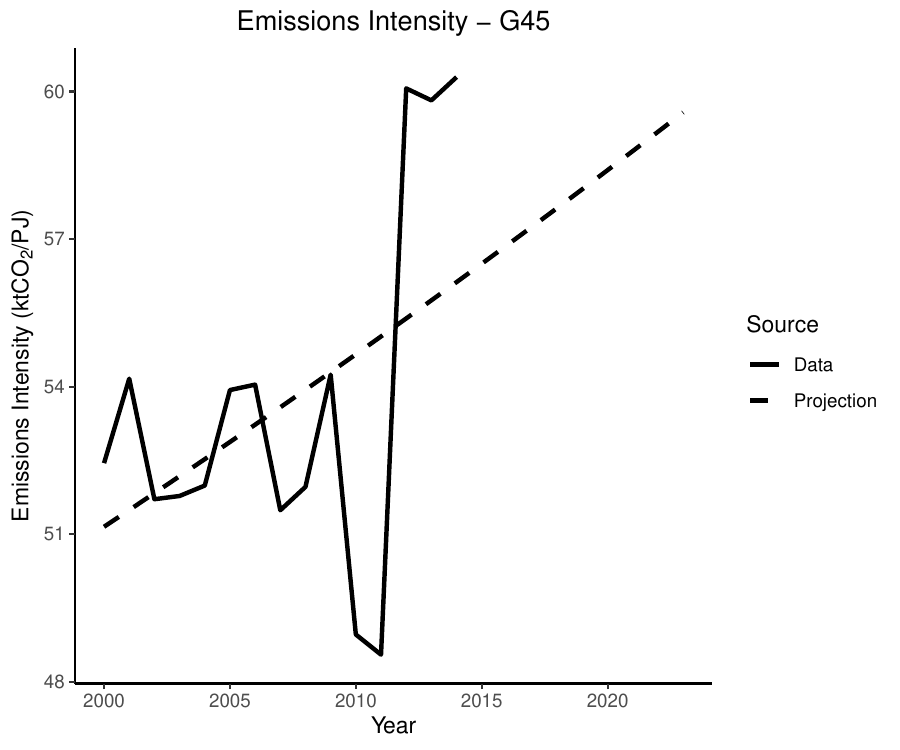}
    \caption{}
    \end{subfigure}
    
    \caption{Projection of Energy Intensity and Emissions Intensity}
    \label{fig:WIOD Intensity Projections}
\end{figure}

\begin{figure}[h!]
\centering
\begin{subfigure}[t]{0.33\textwidth}
        \centering
        \includegraphics[width=\linewidth]{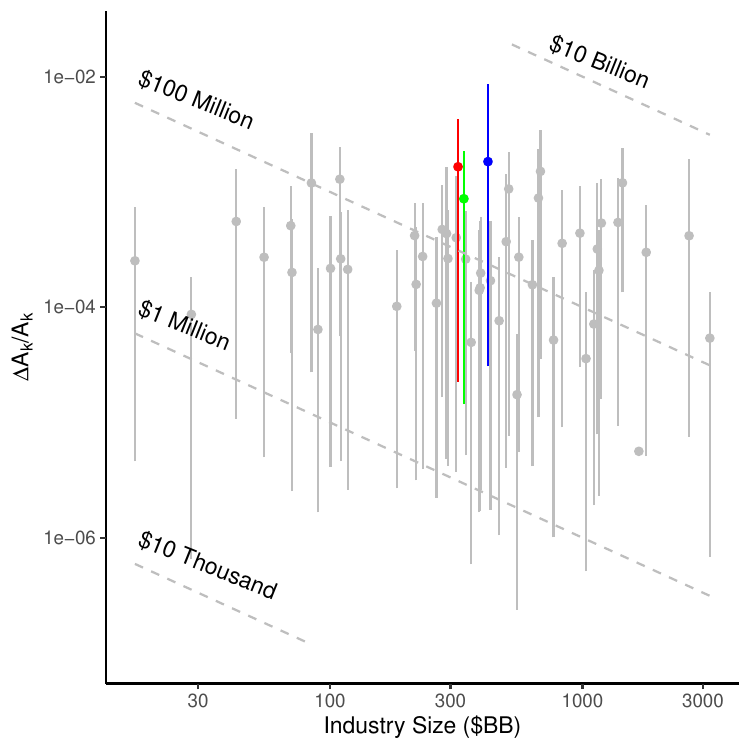}
        \caption{Change in GDP}
    \end{subfigure}%
    ~ 
    \begin{subfigure}[t]{0.33\textwidth}
        \centering
        \includegraphics[width=\linewidth]{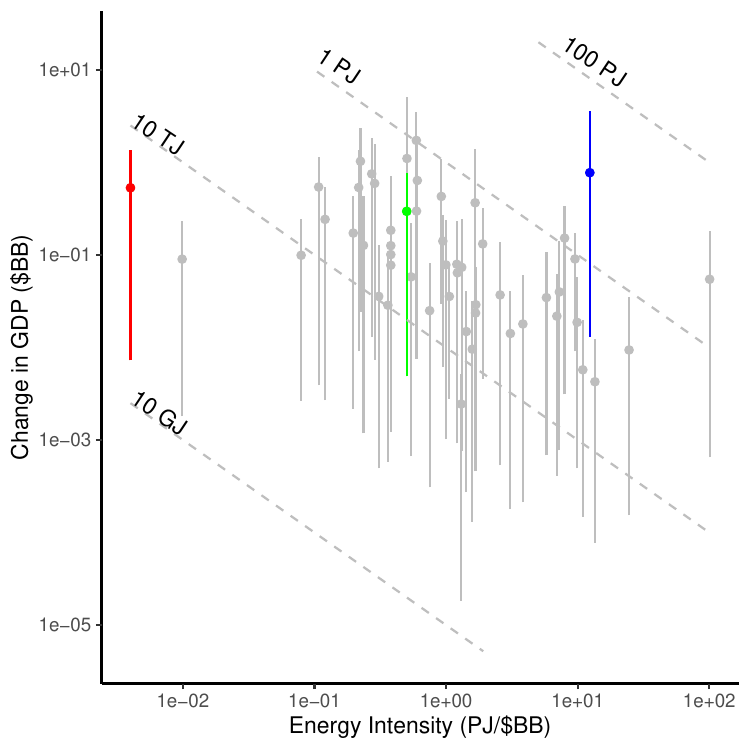}
        \caption{Change in Energy Use }
    \end{subfigure}%
    ~
    \begin{subfigure}[t]{0.33\textwidth}
        \centering
        \includegraphics[width=\linewidth]{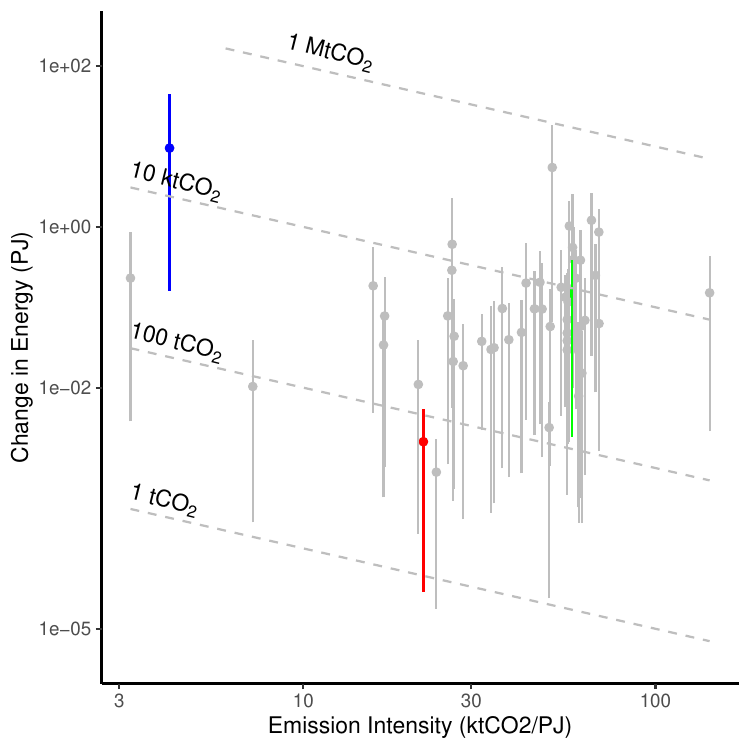}
        \caption{Change in Carbon Emissions}
    \end{subfigure}
    \caption{The Impact of AI on Output, Energy, and CO$_2$ Emissions using projected data. ISIC Rev. 4 Code P85: {\color{blue} Education (Blue)}, ISIC Rev. 4 Code J58: {\color{red} Publishing Activities (Red)}, and ISIC Rev. 4 Code G45: {\color{green} Wholesale and retail trade and repair of motor vehicles and motorcycles (Green).}}
    \label{fig:impact_industry_projected}
\end{figure}

\begin{table}[h!]
\caption{The Impact of AI on Output, Energy, and CO$_2$ Emissions using projected data: Selected Industries}
\label{tab:SI example_industries}
\centering
\scalebox{0.75}{
\begin{tabular}{l|cccc|ccc}
\hline
 \multirow{2}{*}{\textbf{Industry}} & \textbf{Output}  & \textbf{Exposure rates} &\textbf{Energy Intensity} & \textbf{Emissions Intensity}  & $\boldsymbol{\Delta y_k}$ & $\boldsymbol{\Delta E_k}$ & $\boldsymbol{\Delta C_k}$ \\
  & (\$BB) & (\%) &(PJ/\$BB) & (ktCO$_2$/PJ)  & (\$BB) & (PJ) & (ktCO$_2$) \\\hline
 {\color{blue}Education} &   422 &0.183   & 12.367 &4.17 &0.774 & 9.567 & 39.882\\
{\color{red}Publishing Activities}         &   321    & 0.165&0.004& 21.93 & 0.531 & 0.002 &  0.047 \\
{\color{green}Trade and Repair of motor vehicles}         &   339    &0.087 &0.502 & 58.03 & 0.296& 0.149 & 8.621    \\\hline
\end{tabular}
}
\end{table}

\clearpage


\section{Theoretical framework}\label{app:Theoretical framework}
In this paper, we estimate the energy use attributed to AI in the United States economy. Our approach leverages Hulten's Theorem, a fundamental principle in productivity and growth accounting. This Theorem posits that the overall growth of an economy can be expressed as a weighted average of the growth rates of individual firms or sectors, with weights determined by their respective shares in total output. Hulten's Theorem provides a robust method for combining micro-level productivity changes into macro-level growth, highlighting that firms or sectors with larger shares of total output significantly impact overall productivity growth. By applying this principle to the AI sector, we offer a simple framework to better understand its contribution to overall energy consumption relative to its economic output and potential environmental impacts. Here we briefly show how our methodology has its roots in an application of Hulten's Theorem.

Consider a closed and efficient economy composed of multiple sectors. According to Hulten's Theorem, the impact of a marginal change in productivity for any one sector $A_i$ on aggregate output of the economy, $Y$, can be described as

\begin{equation}\label{eq:Hulten}
    d\log Y = \lambda_i d\log A_i
\end{equation}

\noindent where $\lambda_i=\frac{p_iy_i}{\sum\limits_i=1^Np_ic_i}$ is the ratio of sales of industry $i$ relative to aggregate output of the economy, also known as the Domar weight \citep{domar_measurement_1961,hulten_growth_1978}. Equation \ref{eq:task impact} derives from Equation \ref{eq:Hulten} taking discrete differences.

Of course, Hulten's theorem considers marginal changes in productivity, which are unlikely to meaningfully change underlying prices or the allocation of factors of production in the economy. In this way, Hulten's Theorem is a partial equilibrium result. For non-marginal changes in productivity, Hulten's theorem represents an approximation that only holds under Cobb-Douglas production technologies. If production technologies are not Cobb-Douglas, then whether the approximation is biased upwards or downwards depends on the underlying structure of the economy \citep{baqaee_macroeconomic_2019}. Thus, as we address in Table 2 and its corresponding discussion, because we are considering discrete changes in productivity due to AI adoption, it is possible that there is some bias in our results due to holding prices and factors of production fixed. Future work considering a general equilibrium rather than partial equilibrium will be needed to determine if this implies our approximation is an overestimate or an underestimate.

Continuing under the partial equilibrium assumption, we can extend Hulten's Theorem to consider implied impacts on energy use and carbon emissions. For this we apply industry-specific coefficients for energy intensity of output, measured as the ratio of physical energy use to output $\nu_i=\frac{E_i}{p_iy_i}$, and carbon intensity of energy use, measured as the ratio of carbon emissions to physical energy use $\mu_i=\frac{C_i}{E_i}$. Because we are operating within a partial equilibrium framework with no change in prices or factors of production, these coefficients are held constant to the change in productivity. Thus, starting with Equation \ref{eq:Hulten} we can describe the change in energy use and change in carbon emissions from a productivity shock to an industry $i$ as,

\begin{equation}
    d\log E = \nu_i \lambda_i d\log A_i
\end{equation}

\begin{equation}
    d\log C = \mu_i \nu_i \lambda_i d\log A_i
\end{equation}

\noindent These Equations are the foundation for Equations (3) and (4) in the text, again considered in discrete changes.

\end{document}